# The Mu2e Experiment

*Robert H. Bernstein**

*Fermi National Accelerator Laboratory, Batavia, IL, United States*

The Mu2e experiment will search for the charged-lepton flavor violating (CLFV) neutrino-less conversion of a negative muon into an electron in the field of a nucleus. The conversion process results in a monochromatic electron with an energy of 104.97 MeV, slightly below the muon rest mass. The goal of the experiment is to improve the previous upper limit by four orders of magnitude and reach a SES (single event sensitivity) of $3 \times 10^{-17}$ on the conversion rate, a 90% CL of $8 \times 10^{-17}$, and a $5\sigma$ discovery reach at $2 \times 10^{-16}$. The experiment will use a intense pulsed negative muon beam. The pulsed beam is essential to reducing backgrounds. The other essential element is a sophisticated magnetic system composed of three consecutive solenoids that form the muon beam. Mu2e will use an aluminum target and examine $\sim 10^{18}$ stopped muons in 3 years of running. The Mu2e experiment is under design and construction at the Fermilab Muon Campus. The experiment will begin operations in 2022, and will require about 3 years of data-taking. Upgrades to other materials than aluminum are already being planned. This article is written specifically for younger researchers to bridge the gap between conference presentations and detailed design reports, and examines issues not covered in the former without the details of the latter.

**Keywords: particle physics, lepton and lepton-flavor violation, muon, experimental design, beyond standard model searches**

***Correspondence:**
*Robert H. Bernstein*
*rhbob@fnal.gov*

## 1. INTRODUCTION

Mu2e will search for the charged-lepton flavor violating process $\mu^- N \rightarrow e^- N$ (muon-to-electron conversion) by measuring the ratio

$$R_{\mu e} = \frac{\mu^- N \rightarrow e^- N}{\mu^- N \rightarrow \text{all muon captures}} \tag{1}$$

The expected reach, as of this writing, will have a single-event sensitivity of (to one significant digit) $3 \times 10^{-17}$, a 90% CL of $8 \times 10^{-17}$, and a $5\sigma$ discovery sensitivity of $2 \times 10^{-16}$. The experiment will begin operations in 2022. The primary beam will start with the Fermilab Booster, supplying 8 GeV kinetic energy protons on target at 8 kW. Mu2e requires $\approx 3.6 \times 10^{20}$ protons-on-target to meet its goals.

There are many excellent articles motivating the search, such as Calibbi and Signorelli [1], de Gouvêa and Vogel [2], or Marciano et al. [3], among many others. [4] is an indispensable reference for anyone working in the field. An experimental overview of charged lepton flavor violation in leptons across the range of experiments is given in Bernstein and Cooper [5].

Mu2e differs from earlier muon-to-electron conversion experiments in three major ways:

- the intensity of the FNAL muon beam is $\sim$ 10,000 times greater than those of the prior generation of muon-to-electron conversion experiments;
- it uses a novel solenoid system for the formation of the muon beam, which takes advantage of the increased intensity while also providing a beam of momentum and sign-selected muons;
- it uses a pulsed beam with a "long" time between pulses to reduce backgrounds.

We will repeatedly discuss those items and the reader should keep them in mind [1].

This examination of Mu2e provides an opportunity to complement global review articles, conference presentations, and detailed design reports. Our discussion is envisaged as a middle-ground for graduate students and researchers new to the field. Therefore we will examine: (1) experimental considerations: specifically, the most significant backgrounds and how they determine the design of the experiment; (2) the formation of the muon beam and the solenoid system used to make it; (3) the detector and monitoring systems. We focus on aspects of Mu2e not normally covered in general talks, such as the extinction system and stopping target monitor. It relies heavily on the Mu2e Technical Design Report (TDR) of Bartoszek et al. [8]; the experiment has progressed past that design but it serves to document the experiment at a sufficient level for our purposes. It generally adheres to the design presented in the TDR with occasional updates.

## 2. PHYSICS PROCESSES AND EXPERIMENTAL CONSIDERATIONS

Muon-to-electron conversion violates lepton family number, changing a muon to an electron in the field of a nucleus through a coherent interaction with the nucleus. Since no neutrinos are produced, muon-to-electron conversion is not a weak interaction: thus an observation of the process can only come from new physics.

Since the nucleus recoils coherently, the outgoing electron is mono-energetic, with an energy of $E_e = m_\mu - B - E_R$, where $B$ is the binding energy and $E_R$ the recoil energy of the nucleus. A stopped muon near the nucleus falls into a muonic 1 s state quickly, in $\mathcal{O}(10^{-15})$ sec. Czarnecki et al. [9] have calculated that the "conversion energy" in aluminum is 104.973 MeV. It is useful to note that the conversion energy in titanium is 104.394 MeV, practically the same value; if Mu2e chooses to replace its initial aluminum target with titanium the difference in conversion energy will be small enough that changes to the detector will not be required as a result of that shift.

The experiment will measure $R_{\mu e}$ by detecting the conversion electron and measuring its momentum. This single-body state has advantages since there are few other processes that produce electrons near the muon mass. Nonetheless, as is usual in rare process searches there are rare backgrounds. These will set the requirements on the beam structure, the detectors used, and the required momentum resolution on the conversion electron. As an introduction, the backgrounds fall into three general categories:

- Intrinsic backgrounds are produced by the same muons used to measure $R_{\mu e}$ and therefore scale with the number of observed muons, and hence protons-on-target. The most important intrinsic background, from the decay of muons in the stopping target in atomic orbit (DIO), has the same time distribution as the signal. This intrinsic background arises from the Standard Model weak decay of muons which we will discuss in section 2.2. DIOs set the resolution needed in the momentum determination and the need for the best resolution possible is reflected in much of the detector design.
- Beam-related backgrounds are associated with the formation of the muon beam. There are several, but the one that sets the choice of the pulsed beam is radiative pion capture (RPC). This background is the primary reason for the experiment's pulsed beam structure. The pulsed beam allows the experiment to (1) suppresses the RPC background that limited earlier experiments and (2) take advantage of the $\times 10,000$ increase in muon flux without being overwhelmed by background or debris from the initial proton beam interaction.
- Cosmic ray backgrounds arise from cosmic ray interactions and/or decays occurring in or near the detector and scale with live time. The suppression of this background will require a large and hermetic veto system that can also survive the high neutron flux arising from the collisions of protons in the production target.

The muons are captured in an aluminum stopping target. Measurements of the muon lifetime [10, 11] report an 864 ns lifetime for a muon in an aluminum orbit compared to the 2.2 $\mu$s free lifetime; applying $1/\Gamma = 1/\Gamma_{\text{capture}} + 1/\Gamma_{\text{free}}$ tells us that about 40% of the muons decay-in-orbit and 60% are captured by the nucleus. The Mu2e stopping target is discussed in section 4.3. Although we have not yet discussed the detectors, it is useful to note that the Mu2e detectors are annular: the muon beam and stopping target are centered along the magnetic axis of the solenoid and the detectors are downstream of it, arranged in annuli. This reduces background and activity in the detector, for reasons we will discuss in the appropriate sections.

Most discussion of muon-to-electron conversion refer to the $Z$ of the nucleus. In fact the rate depends on both $Z$ and $A$. The $Z$ dependence comes from photonic contributions to muon-to-electron conversion, first calculated in Feinberg et al. [12]; the coherent rate goes as $Z^5$ and normalizing to the total capture rate reduces the dependence to $Z^4$. These are particularly interesting for SUSY models; the same magnetic dipole operator that appears in Mu2e for SUSY loops appears in $\mu \rightarrow e\gamma$, but in $\mu \rightarrow e\gamma$ the photon is real. SUSY of course is not the only possible source of charged lepton flavor violation and we need to be more general. Incoherent photonic contributions have no such $Z^5$ enhancement. Non-photonic contributions depend on $A$ rather than $Z$. Ultimately in a given model one needs the quark content and quark couplings, often translated into nuclear form factors.

[1] The COMET experiment at J-PARC is similar in many regards to Mu2e but unfortunately we do not have space to discuss it here; details can be found in Kuno [6]. The DeeMe experiment at J-PARC plans to use a quite different technique to improve the measurements by a single order-of-magnitude; Mu2e and COMET propose a $10^4$ improvement, but the reader is referred to Aoki et al. [7] for a presentation of the DeeMe method. The construction of both experiments is steadily progressing at this writing.

## 2.1. The $\Delta L = 2$ Process $\mu^- N \rightarrow e^+ N'$

There is an increasing interest in the $\Delta L = 2$ process of $\mu^- N \rightarrow e^+ N'$ with $\Delta Z = 2$. This process can occur through transitions to the ground state of the final nucleus, or through transitions to an excited final state. If the final state is the ground state of the excited nucleus, the positron is mono-energetic; transitions to an excited final state are not, and it is far more difficult to observe a signal. Radiative muon capture, $\mu^- N \rightarrow \gamma N'$ (section 2.4.1) with a subsequent conversion, is particularly problematic here. It is an intrinsic background since it is generated by the same muons captured for the $R_{\mu e}$ measurement, and the spectrum in the relevant region is poorly known. See Berryman et al. [13] or Geib et al. [14] for a current assessment of the theory and Bernstein and Cooper [5] for the experimental history.

## 2.2. Decay-in-Orbit Background

Muonic weak decay, $\mu^- \rightarrow e^- \bar{\nu}_e \nu_\mu$ is well-understood and is covered in many textbooks, such as Commins [15]. We are concerned with the momentum spectrum for the decay of a muon. For free muons, the spectrum is commonly called the Michel spectrum after Michel [16]. The Michel spectrum has an endpoint at

$$E_{\max} = \frac{m_\mu^2 + m_e^2}{2m_\mu} = 52.8 \text{ MeV} \qquad (2)$$

which one can derive from four-momentum conservation, with the assumption of negligible neutrino mass. This is so far below the endpoint of muon-to-electron conversions near the 105.66 MeV/c$^2$ muon mass that with even modest resolution there is practically no background from free muon decay. However, in the decay of a bound muon (DIO, or decay-in-orbit) the spectrum is altered because the outgoing electron can exchange a photon with the nucleus. The recoil of the electron off the nucleus can make the electron's final energy equal to the conversion energy. This is not difficult to understand. The final state of muon-to-electron conversion is an electron recoiling against a nucleus. The final state of a decay-in-orbit event is an electron and two neutrinos recoiling against a nucleus. At the decay-in-orbit endpoint where the (massless) neutrino energy is zero, the two final states are the same and have the same electron energy.

We can understand the general form of the spectrum near the endpoint by a phase space calculation, known as Sargent's rule, which tells us it behaves as $(E_{\text{conv}} - E)^5$ (also of great use in understanding direct searches for non-zero neutrino mass; see [17]). Czarnecki et al. [9] have calculated the spectrum, followed by a calculation of the $\approx$ 10% radiative corrections in Czarnecki et al. [18]. We reproduce the spectrum from Czarnecki et al. [9] in **Figure 1** (using the earlier paper because it contains plots useful for our purposes). Examining the right-hand side of **Figure 1** provides a quick estimate of the required resolution. A measurement $\mathcal{O}(10^{-17})$ requires at least $10^{17}$ muons; the graph shows we could expect $\sim$ 1 event within an MeV of the muon-to-electron conversion endpoint. As a rough estimate, the momentum resolution must therefore be $\approx$ 1 MeV/c or less; a detailed simulation tells us the experiment requires a resolution of about 180 keV/c. This value is a Gaussian characterization of the core of the resolution function, but as discussed later the experiment requires the tails of the resolution function are well-understood. These two challenging goals, good core resolution with small tails, determine much of the detector design.

Before discussing individual backgrounds we give the overall time structure of the beam. Mu2e will use a "pulsed" beam as shown in **Figure 2**. The period is 1,695 ns, the revolution period in the Fermilab Delivery Ring (formerly the Fermilab Recycler). The pulse shown is Gaussian for clarity, but the width of $\approx$ 125 ns is approximately correct. Protons striking the target make pions. Those pions will be allowed to decay and the resultant muons will be made into the Mu2e muon beam. **Figure 2** shows the arrival time of the pions at the stopping target, along with the arrival time of muons. The muons that are captured by the stopping target decay with their 864 ns lifetime. A "selection window," "measurement period," or "live time" (all phrases used to refer to the same concept) starts about 700 ns after the pulse. We can see two important concepts from this Figure. First, many pions arrive early, peaking at about 200 ns with a long tail. Second, since the muon lifetime is 864 ns, about half the 1,695 ns repetition rate, the loss from beginning the selection window at 700 ns is acceptable (we will discuss the reason for this wait in the section 2.4).

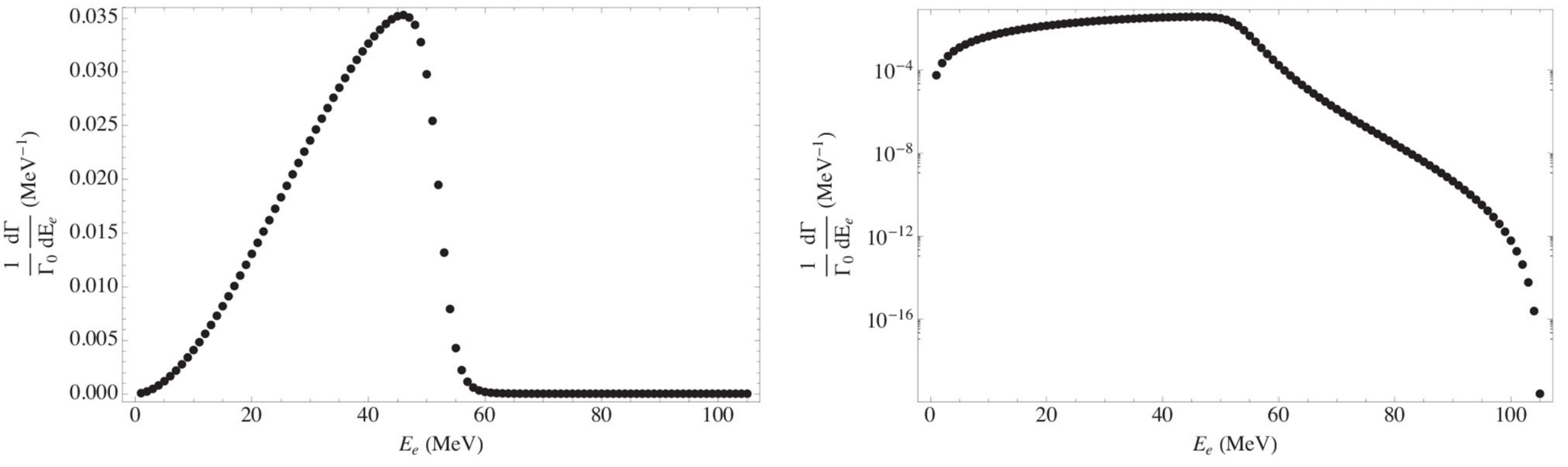

**FIGURE 1 |** Decay-in-orbit electron spectrum in aluminum near the endpoint, normalized to the free-muon decay rate $\Gamma_O$. The Figure is taken from Czarnecki et al. [9]. The left-hand plot is on a linear scale, the right a logarithmic one.

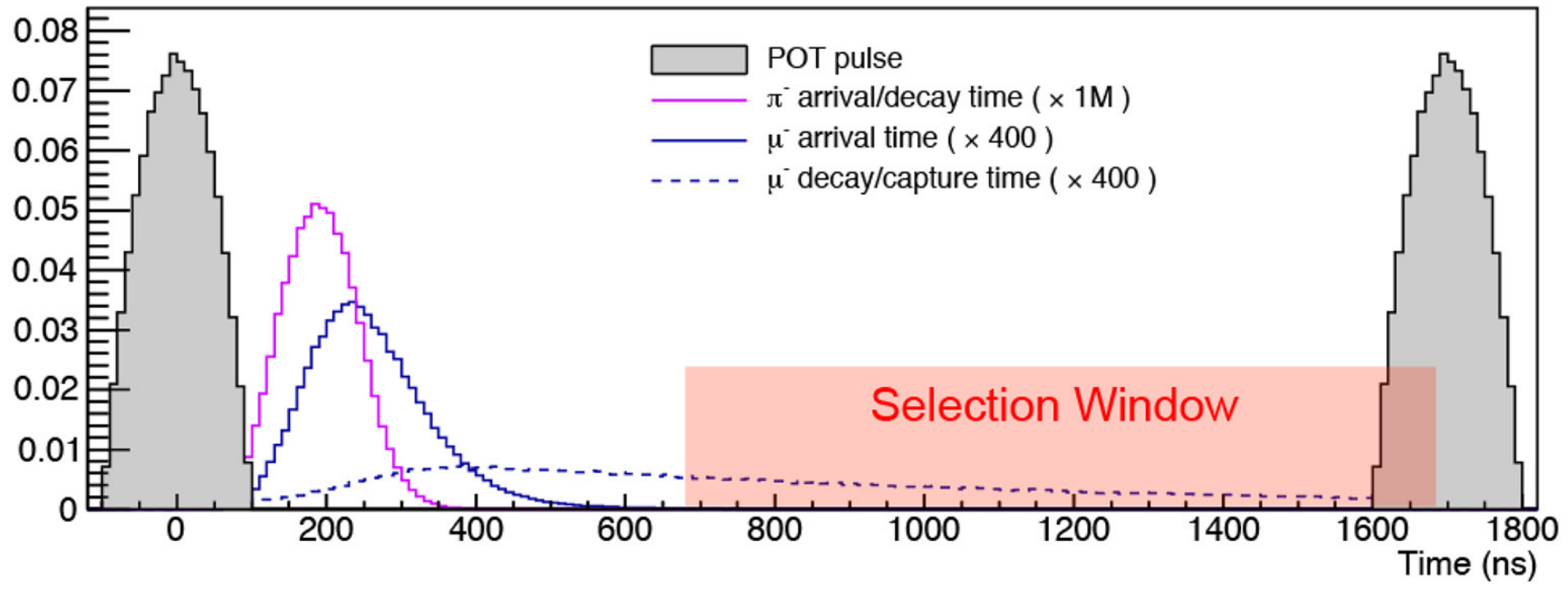

**FIGURE 2 |** The Mu2e beam timing. A pulse with $\approx 3.9 \times 10^7$, 8 GeV kinetic energy protons arrives every 1,695 ns. The arrival time distribution for pions that arrive at the detector solenoid (greatly reduced by the Mu2e Solenoid system) are shown, along with the arrival time for muons and the decay or capture time (both with an 864 ns lifetime). The "Selection window" is the period of time for which Mu2e will analyze data; the live gate will take data as early as about 500 ns, with the final accepted region determined through analysis. The pulse shape is more complicated than the idealized form shown here. See Bartoszek et al. [8] for more information.

## 2.3. Beam Flash

What is not shown in **Figure 2** is the "beam flash" that determines many aspects of both Mu2e and potential upgrades; the beam flash is also the likely limitation of the entire method. In addition to charged pions, the proton collisions make neutral pions, and $\pi^o \rightarrow \gamma\gamma$ produces photons that can convert in the primary target. The electrons from these conversions can be transmitted to the Mu2e stopping target and detector. This "flash" overwhelms the detector and make it impossible to find a signal during the flash's time period. Muons that stop in the target and undergo capture also produce protons, neutrons, and photons: $\mu^- + {}^{27}\mathrm{Al}_{13} \rightarrow \nu_\mu + X + ap + bn + c\gamma$, and the outgoing particles can produce accidental activity and radiation damage (see [10] for typical $p$, $n$, and $\gamma$ multiplicities). The flash is over after a few hundred ns, so the analyzed data will not see them if the analysis period begins at 700 ns. However, the flash provides both a technical and a physics limit. First, as the intensity increases, the radiation damage from the flash increases, leading to shorter lifetimes for detector elements. Second, the physics limit on the start of the measurement window comes from the lifetime of a muonic atom. The lifetime of a muon in an atom will decrease with $Z$: there are more nucleons to interact with, and the Bohr radius decreases, increasing the overlap with the nucleus. For reasons we will see later, it will be interesting to go to higher $Z$ materials to probe possible signals.

Unfortunately we cannot go to arbitrarily high $Z$. In general, a muon arriving at $t'$ decays as $e^{-(t-t')/\Gamma\times\tau(A,Z)}$, but the muon beam itself, which determines the distributions of $\Gamma$ and $t'$, is the same regardless of the $Z$ of the target ($\Gamma$ is $E/m$ and $\tau(A,Z)$ is the lifetime in the atomic state). **Figure 2** was made for an aluminum target; the dashed $\mu^-$ decay/capture curve represents the arrival time of the muons combined with the 864 ns aluminum lifetime. The decays in aluminum are well separated from the flash. With muonic gold, an excellent candidate material, the lifetime is only about 74 ns. In moving from gold to aluminum $\tau(A,Z)$ changes from 864 to 74 ns, the exponential drops more quickly, and as a result the decays occur mostly during the time pions are still arriving. At these early times the physics backgrounds are large and the accidental activity high, and no useful measurement can be made. No practical way out of these problems with this technique for forming muon beams exists. Exploring high $Z$ materials will require a new method, which we will discuss in the context of upgrades.

The reader is asked to keep the 700 ns number and **Figure 2** in mind throughout. The reader should also be aware that this value for the beginning of the measurement period will not be set by hardware or a trigger and the final value will be determined from the data, and in fact with modern analysis methods the experiment is more likely to use probability density functions than a hard cut.

## 2.4. Radiative Pion Capture

Radiative Pion Capture (RPC) is the process $\pi^- N \rightarrow \gamma N'$ where $N'$ is an excited nuclear state; RPC takes place with a probability $\approx 2 \times 10^{-2}$ (see [19]). The spectrum peaks around 110-120 MeV and an asymmetric conversion can yield an electron at the conversion energy. This conversion can occur either by the photon externally converting in the material of the stopping target or undergoing an internal conversion, $\pi^- N \rightarrow e^+ e^- N'$. Kroll and Wada [20] have calculated the internal conversion fraction for $\pi^- p$ radiative pion capture, and the experiments so far have all assumed the same ratio for captures on complex nuclei. For Mu2e, by numerical coincidence, the internal and external conversion probabilities for the Mu2e target and geometry are approximately equal (the probability of conversion in the stopping target as the photon propagates through the Mu2e stopping target geometry is about the same as the internal conversion fraction). For completeness, we note that the number of $e^-$ produced in association with external conversion is greater than the number of $e^+$: the outgoing photon can Compton scatter, which yields $e^-$ but not $e^+$. This will also be true for Radiative Muon Capture, discussed in section 2.4.1.

The background from RPC electrons is perhaps the single biggest determinant of the proton beam time structure. Pions born in the initial proton collisions with the "production" (as opposed to the "stopping" target) are used to make muons. The best existing $\mu^- N \rightarrow e^- N$ search, by the SINDRUM-II experiment [21], was performed at the Paul Scherrer Institut (PSI) and measured $R_{\mu e} < 7 \times 10^{-13}$ at 90% CL. The PSI beam structure provides a pulse of muons ∼300 ps long every 19.75 ns. This time between pulses is about the same as the pion lifetime and after the increase in pion lifetime from $\gamma$ and the long muon lifetime the muon decays are effectively constant. SINDRUM-II required that fewer than 1 out of $10^9$ pions reached their stopping target or electrons from RPC would overwhelm a signal. SINDRUM-II used a degrader to remove pions, along with veto counters to remove out-of-time pions after the beam pulse. Mu2e plans to reach $10^4$ further in its search, and the correspondingly $\mathcal{O}(10^4)$ more intense beam needed makes the SINDRUM-II method unworkable.

The 19.75 ns beam structure of PSI will be replaced with a 1,695 ns repetition rate: therefore, as we saw in **Figure 2**, the muon beam arriving at the stopping target will be cleanly separated in time from the flash and the majority of RPCs. The Mu2e pulsed beam, combined with the delay discussed earlier, is the key to taking advantage of the increased intensity; without it the experiment would be limited by a combination of the activity from the beam flash and how well it could measure the RPC rate and spectrum.

Although **Figure 2** seems to show a negligible number of pions arriving inside the "selection window," recall that Mu2e is measuring a process at the $10^{-17}$ level. The tail of the pion arrival time distribution, negligible on the scale of this Figure, is large for the needs of the experiment. A detailed beam simulation tells us the number of pions is suppressed by $\mathcal{O}(10^{11})$ if the measurement period begins at about 700 ns, reflecting a combination of the time to transit the beamline and the short pion lifetime. This delay then yields an acceptably small RPC background. This is the reason for the pulsed beam: the experiment will wait until there are few enough pions to reach a manageable background. Unfortunately, forming a pulsed proton beam is not a perfect process. Sometimes protons "outside" the pulse can be transmitted. If these protons strike the production target in-between beam pulses, the produced pions can "restart the clock" and evade the $\mathcal{O}(10^{11})$ suppression. Mu2e has calculated that the in-time to out-of-time ratio needs to be better than $10^{-10}$, assuming a flat distribution for the out-of-time beam; this ratio is called "extinction" in Mu2e. Note this extinction is a different concept from the delayed measurement period, but is related. The common motivation is the same: keep pions out of the detector during the live time.

#### 2.4.1. Radiative Muon Capture

Radiative Muon Capture (RMC) is analogous to radiative pion capture, $\mu^- N \rightarrow \gamma N' \nu_\mu$. However, the spectrum is considerably softer than the RPC spectrum and the photons (specifically their converted electrons) are not at high enough energy to produce a significant background. The RMC background can distort the DIO spectrum since they overlap in the 80–100 MeV/c range. This is a problem because the experiment will check the theory prediction for decays-in-orbit at lower momenta before using it for the endpoint; if the RMC spectrum is large and not well-known, the DIO prediction will not agree with data and the extrapolation to the endpoint might not be trusted as a result. The softer spectrum does provide a potentially significant background for $\mu^- \rightarrow e^+$ conversion, as mentioned in the discussion of section 2.1. The RMC spectrum in the interesting region is poorly known, and one of the experimental challenges will be to determine it.

The theoretical calculations are in need of improvement. The last calculations used the "closure approximation," which replaces a sum over nuclear states by transitions to an average energy [22]. The interested reader is referred to Bergbusch et al. [19] for the last measurement. The data are somewhat old and the statistics limited in the region needed by Mu2e.

### 2.5. Cosmic Ray Background

A through-going cosmic ray can strike the Mu2e stopping target, knock out an electron with the conversion energy, and exit the detector with no other trace. Such an electron is indistinguishable from the signal since it comes from the stopping target and has the right energy. Mu2e estimates there would be approximately one such electron per day, which would yield $\mathcal{O}$ (1,000) events over the expected run. Cosmic rays can also decay while in the detector volume or interact with the detector elements or other parts of the apparatus.

Suppressing these backgrounds calls for a nearly hermetic cosmic ray veto; limiting the background to an acceptable level implies the efficiency of the veto system must be ≈ 99.99%.

### 2.6. Other Backgrounds

Backgrounds from muon and pion decay-in-flight will be negligible. One can make a physics argument that shows why this is the case. We will see Mu2e uses a graded solenoidal field in section 3.1. The field is graded gently enough to use the adiabatic invariance of the flux. Assuming $p_\perp^2/B$ is constant we obtain: [23]

$$v_\parallel^2 = v_o^2 - v_{\perp 0}^2 \frac{B(z)}{B_o} \tag{3}$$

where we imagine the magnetic field points along the $z$-axis and then define $v_\parallel$ and $v_\perp$ with respect to $z$. $B_o$ and $v_o$ are the initial field and velocity.

Assume a linear decrease of the field; then propagating from $z$ to $z + \Delta z$ with initial angle to the field $\theta_o$, one can derive the arrival time (assuming a small time has elapsed since passing PhD qualifiers, or the ability to use [24]):

$$t = 2\frac{\Delta z}{v\sin^2\theta_o}\left(-\cos\theta_o\left(\frac{B_o}{\Delta B}\right) + \sqrt{\cos^2\theta_o\left(\frac{B_o}{\Delta B}\right)^2 + \sin^2\theta_o\left(\frac{B_o}{\Delta B}\right)}\right) \quad (4)$$

By demanding the decaying $\pi$ or $\mu$ to have sufficient energy to produce a signal electron, relativistic kinematics can be used to show they will normally arrive no later than 500 ns after their birth, 200 ns before the beginning of the signal measurement period. This is only a qualitative argument and detailed simulations are required (especially of the proton beam pulse time distribution) but this simple argument indicates why muon and pion decay-in-flight are negligible backgrounds.

A potentially significant background is from antiprotons produced in the initial proton collision in the production target. Antiprotons tend to have smaller kinetic energies and can therefore arrive at the stopping target any time between beam pulses. Since the antiproton lifetime is effectively (if not) infinite, they evade the 700 ns wait time that suppresses pions. The antiprotons can then annihilate in the stopping target and produce, at the nominal exposure, thousands of background events. A system of thin windows annihilates the antiprotons before they can reach the stopping target with an acceptable loss of muons.

The design of the antiproton suppression windows is complicated by the poor knowledge of the relevant differential cross-sections. The Mu2e beam energy is 8 GeV and, as we will see, Mu2e mostly looks at backwards production. This presents two difficulties. First, there is a threshold for antiproton production, and precise measurements near thresholds are difficult. In this case, the threshold arises because the production of an antiproton must conserve both charge and baryon number. Therefore, the *pp* collision must have enough at least energy for the $p + p \rightarrow p + p + (\bar{p} + p)$ process. Energy-momentum conservation gives a threshold of 5.2 GeV (about 4.1 GeV with the extra energy from Fermi motion). The 8 GeV beam is near enough to that threshold so that the cross-section changes relatively rapidly with energy. Many measurements have been performed for cosmic-ray produced antiprotons, but those are mostly for *pp* collisions (for example, [25]). These cannot be used directly: the differential cross-section is altered on the heavy ($W$) production target. A series of measurements were made ([26] and references therein) on heavy targets, but the data are limited: there are no data past about 119 degrees from the forward (initial proton) direction, and no real model with which to perform an extrapolation. While Mu2e believes this background is under control, upgrades to the experiment that run at lower energy ($\sim$1–3 GeV, for example, at FNAL's PIP–II) below $\bar{p}$ production threshold would eliminate the problem [27].

# 3. THE Mu2e PRIMARY BEAM, SOLENOIDS, AND MUON BEAM

## 3.1. Overview

Mu2e will use the Fermilab Delivery Ring to create a primary proton beam at 8 GeV kinetic energy at a power of about 8 kW. The pulse width will be about 250 ns FWHM with pulse spacing 1,695 ns apart. There is a complicated macrostructure to allow delivery of beam to the FNAL neutrino program but we will not examine that here; the macroscopic duty factor for Mu2e is about 30%. The overall timing of a typical cycle was shown in **Figure 2**, with more discussion in, for example, section 2.4.

The Mu2e muon beam is then formed through the Mu2e solenoid system. The solenoid system is the most innovative, technically challenging, and singularly essential part of the experiment. First, the idea does not come from Mu2e: the three-part solenoid system of Mu2e was invented by Abadjev et al. [28]. One can see from examining that document that in some sense very little has changed: the core of the ideas is there. A sketch of the Mu2e solenoids appears in **Figure 3**; see Bartoszek et al. [8] for more information.

The solenoids perform several critical functions for the Mu2e experiment. Their magnetic fields are used to efficiently collect and transport muons from the production target to the muon stopping target while minimizing the transmission of other particles. Electrons are transported from the stopping target to detector elements where a uniform and precisely measured magnetic field is used to measure the momentum of electrons. The magnetic field values range from a peak of 4.6 T at the upstream end of the first solenoid to 1 T at the downstream end of the last solenoid. In between is a complex field configuration consisting of graded fields with a final, nearly uniform field region over the detector region. The field of each region was designed to satisfy a very specific set of criteria. The system (also used in the COMET experiment with variations) consists of three sections of superconducting solenoids: Production (called "Capture" by COMET), Transport, and Detector.

### 3.1.1. Production Solenoid

The Production Solenoid contains a radiatively cooled tungsten production target. The proton beam comes (in **Figure 3**) from the right, which defines the "forward" direction. The muon beam exits away from the initial beam direction (in fact, the beam enters at $17^o$ to the axis of the solenoid system). The Mu2e field at the target is about 4.3 T with a maximum of about 4.6 T. Backwards-going pions decay into muons. Some forwards-going muons are reflected in the magnetic field or scatter and travel backwards, but this is a relatively small fraction. Using Equation (3) and substituting these values tells us that particles produced at less than about $73^o$ to the magnetic axis are not reflected in the Production Solenoid back toward the Transport Solenoid.

The reason for using this "backwards" scheme is simple. An overwhelming flux of particles (including neutrons, photons, and electrons or positrons from photon conversion) are produced in the forward direction, and the leftover incoming proton beam must be absorbed. Mu2e will see $\approx 3.9 \times 10^7$ protons per pulse. The detector needs to be capable of isolating a process at the $10^{-17}$ level and simply could not withstand the resultant flux. The solenoidal field focuses and the gradient ($\approx$ 0.28 T/m along the magnetic axis) directs outgoing muons (and undecayed pions and antiprotons) into the second section, the transport solenoid. Nonetheless, the overall efficiency is quite high: approximately 0.005 muons/proton-on-target reach the aluminum stopping target; the muons have a peak momentum of about 35 MeV and the number of stopped muons/proton-on-target is about 0.002. The 0.002/0.005 = 40% stopping fraction was the result of an optimization: making the target thicker to stop all the muons would increase the multiple scattering and energy loss of outgoing conversion electrons and ultimately increase the backgrounds. **Figure 5** presents the momentum distribution and rates at the stopping target.

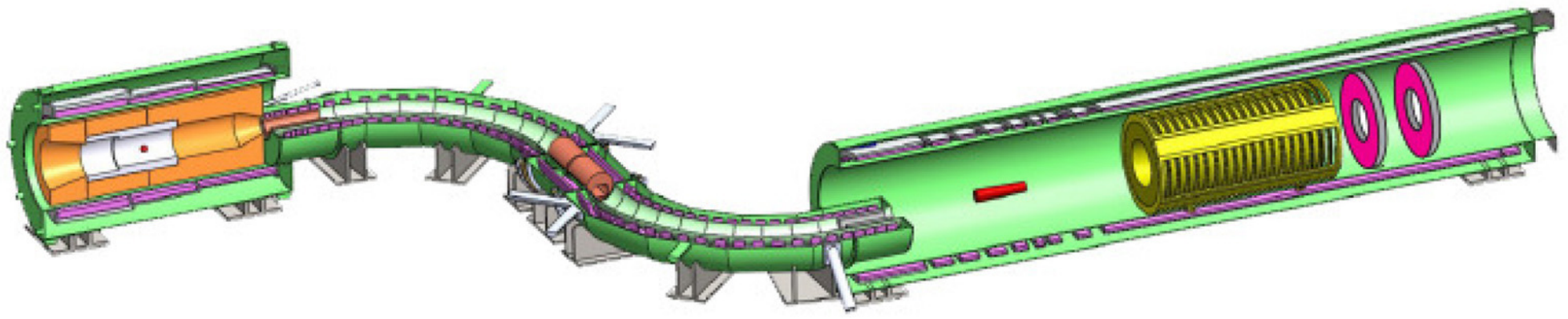

**FIGURE 3 |** The Mu2e Solenoid system, also showing the tracking detector, calorimeter, and stopping target. From Bartoszek et al. [8].

The field gradients throughout the system have been carefully designed. The field generally drops as muons move along the system; this negative gradient directs muons from the Production Solenoid to the Detector Solenoid. Any positive gradients have to be carefully understood; such gradients can appear in transition regions between coils, for example, and can result in "trapped" particles. These trapped particles can be a problem: for example, a pion or muon at sufficiently high momentum can be trapped and decay late. Equation (4) will not apply for these—the adiabatic invariance of the flux is not valid for these small traps—and the consequent "late arriving" backgrounds can be problematic.

The Production Solenoid has a port for the transmission of uninteracted beam to an absorber, as well as a port for scattered protons at $\sim$ 4 GeV to proceed to an "extinction monitor" used to measure the out-of-pulse beam, needed for the reasons discussed in section 4.2.

One of the most difficult problems in Mu2e is the design of the heat and radiation shield ("HRS") for the Production Solenoid. The Mu2e magnets use $\approx$ 75 km of NbTi superconducting cable stabilized with high conductivity aluminum. The beam, with 8 kW of power, would drive the superconductors normal without a cooling system. Mu2e will use indirect cooling with liquid helium. We cannot possibly cover the details, and choose only to define a few terms that appear frequently and are not normally discussed or defined in general articles for experimenters. Radiation damage to the superconductor, the aluminum "stabilizer" holding the superconductor, and to the superconducting coil are all issues; here we mention two issues for the non-superconducting aluminum stabilizer.

Why do superconducting magnets use stabilizers? One reason is mechanical stability. If a superconducting current-carrying cable with current density $J$ moves a distance $\Delta x$ in the magnetic field $B$, the work done per unit volume is just $JB\Delta x$. If some of that work is dissipated into heat (through, for example, friction or impact against another wire) it can drive the superconductor normal. Shifts of order tens of microns are sufficient for quenching. Another reason is "flux jumps": without being too technical, in Type II superconductors magnetic fields induce screening currents inside the superconductor; any change in the screening current allows flux to move into the superconductor; the motion of flux dissipates energy; dissipation of energy raises the temperature; the critical current density for superconductivity falls as the temperature rises, changing the screening current in a positive feedback loop until the magnet quenches. Superconductors are poor conductors when normal, and the stabilizer provides a temporary lower-resistivity path to allow heat to escape, making it possible for the superconductor to recover [29].

Radiation can displace atoms from the stabilizer lattice, measured in "displacements per atom," or DPAs. Impurities such as dissolved interstitial oxygen, carbon, nitrogen, or hydrogen are a particular concern; such impurities can serve as scattering centers for electrons, making the flux-jump stabilization less effective. Such degradation is quantified in the "ratio of residual resistivity" or RRR. Considerable effort has gone into the design of the HRS and cooling system, with a more complete description of this and many other issues (such as quench protection) in the solenoid system in Bartoszek et al. [8]. A standard method of repairing such damage is "annealing." Annealing brings the system to higher temperature, allowing thermal motion to "repair" the lattice, and then cooling once again (we will see the same idea in repair of the germanium detector in section 4.4).

### 3.1.2. Transport Solenoid

The Transport Solenoid's "S" shape serves three major functions. First, photons are not transmitted. Next, and this would be true for a straight solenoid as well, particles with too large a momentum (radius of gyration) hit the walls of the solenoid and are not transmitted, with the radius of gyration given by

$$r = \frac{P_\perp}{0.3B} \tag{5}$$

Third, positive muons will not be captured by the nuclei of the stopping target and therefore should not enter the detector region where they would only produce increased accidental activity or backgrounds. The "S" shape takes advantage of particle drifts in the solenoidal field as shown in Jackson [23] to remove positive muons. In Mu2e's case, $\mu^-$ and $\mu^+$ drift vertically in opposite directions. A central collimator covering "half" the aperture blocks the positive $\mu^+$ and transmits the desired $\mu^-$. **Figure 4** is a representation of how the arrangement works. Numerically, the vertical displacement $D$ midway through the TS is given by (for pathlength $s$, radius $R$ with $s/R = \pi/2$ and $B$ in Tesla, momenta in GeV/c, and the components of momentum parallel and perpendicular to the magnetic axis given by $P_\parallel$ and $P_\perp$):

$$D = -\frac{\pi}{2}\frac{e}{Q}\frac{1}{0.3B}\frac{P_\parallel^2 + \frac{1}{2}P_\perp^2}{P_\parallel} \quad (6)$$

and the sign of $Q$ in Equation (6) reflects the separation of positive and negative particles. It is interesting to note that for the forward-moving beam, $D$ is approximately proportional to $P_\parallel$, so higher momentum particles have greater deflections. The second half of the "S" shaped Transport Solenoid brings the beam back to the nominal axis (and provides additional length for pions to decay, suppressing the RPC background).

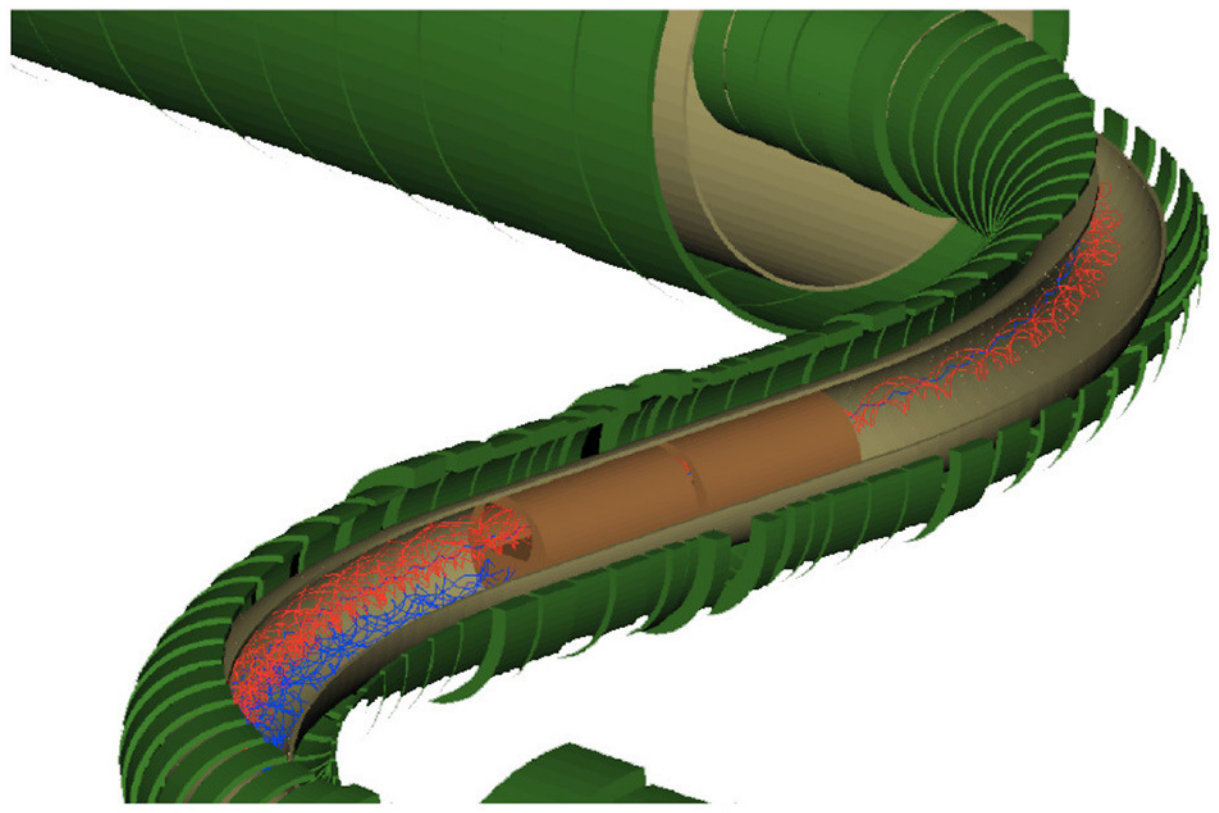

**FIGURE 4 |** Charge selection in the central Mu2e collimator. Red $\mu^-$ are bent upwards; $\mu^+$ are bent down and strike the collimator in this Figure; the experiment will deflect in the opposite direction from what is shown here. From Bartoszek et al. [8].

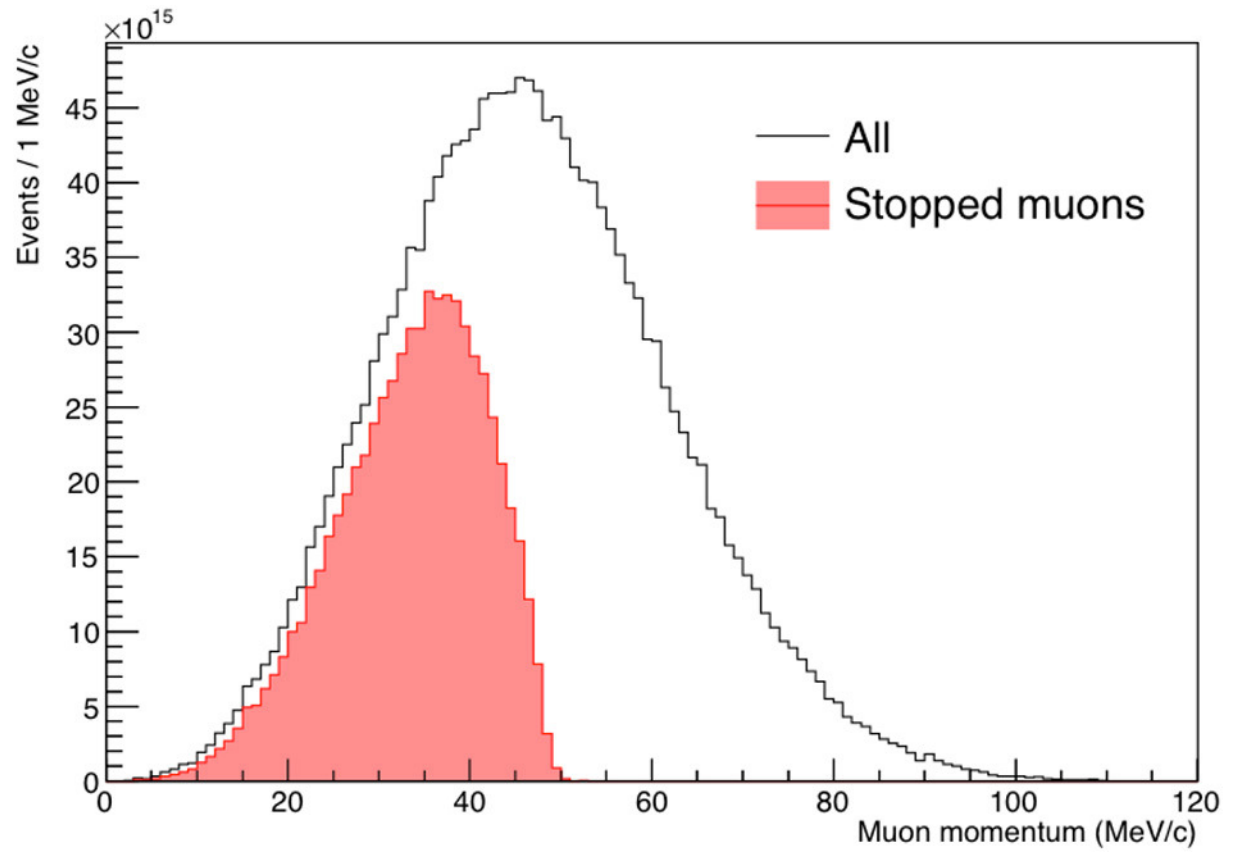

**FIGURE 5 |** Momentum distribution of all muons reaching the stopping target together with those that do stop, normalized to the standard exposure of $3.6 \times 10^{20}$ protons-on-target. From the Mu2e Collaboration.

The transport solenoid, starting at about 2.5T with a gradient along its $\sim$ 12 m arc length, yields a muon beam with a peak momentum around 45 MeV/c at the entrance to the Detector Solenoid. The momentum distribution of muons reaching the stopping target is shown in **Figure 5**. Mu2e will only stop about half the muons; material for stopping muons will increase multiple scattering and energy loss for the conversion electrons, and both of these stochastic processes will worsen the signal/background separation. The momentum distribution of muons that do stop is also shown in **Figure 5**.

### 3.1.3. Detector Solenoid

The Detector Solenoid contains the stopping target and the detector system. The muon beam passes through the stopping target and about half stop; the rest proceeds into a beam dump still in vacuum. A small, thin window at the downstream end allows the experiment to detect X-rays produced when muons stop and cascade to a muonic 1 s state, as described in section 4.4. It is worth noting that this solenoid has no bend, unlike COMET; therefore both $e^+$ and $e^-$ can be observed simultaneously. The disadvantage of the straight detector solenoid is that the detector is exposed to the beam flash and the remnant muon beam. With COMET's bend the detector does not need to be annular: since the momentum of transmitted particles can be selected with the bend, neither the low momentum muons nor the bulk of the decay-in-orbit spectrum will reach the detector. The advantage of the straight solenoid is that the detector geometry is charge-symmetric. Since photons from radiative pion and radiative muon capture produce equal numbers of electrons and positrons, both of those backgrounds can be measured *in situ*. The charge symmetry also enables a simultaneous measurement of the $\Delta L = 2$ process $\mu^- N \to e^+ N$. Radiative muon capture is a potentially significant background here because the spectrum in the relevant region is poorly known.

The muon beam dump needs to be inside the solenoid's vacuum. If the remnant beam were to pass through a window, muons would lose energy and stop in the window, and send particles back into the tracker.

The Detector Solenoid is graded from about 2.5 T at the transition from the Transport Solenoid down to 1 T in the detector region. The gradient after the stopping target has an additional purpose: it causes the conversion electrons to be pitched forward into the acceptance of the tracker (the field in the detector region itself is relatively constant).

## 3.2. Antiproton Background and the Muon Beam

It is worth noting that $\bar{p}$'s can be produced in the production target since the beam energy is above the threshold for $\bar{p}$ production and these negatively charged particles can pass through the sign-selecting collimators. Antiprotons tend to have a higher momentum than muons, have a larger radius of gyration, and not be accepted by the solenoid. Nonetheless, internal Mu2e simulations show thousands of electrons in the signal window would appear from antiproton annihilations in the stopping target, yielding backgrounds unacceptably close to the existing limits. Antiprotons that annihilate too far down the beamline "regenerate" pions and therefore can increase the radiative pion capture background. A tradeoff among maximizing the stopped muon yield, minimizing the regeneration of pions, and using the solenoid acceptance and absorbers led to the final system. Antiprotons tend to have such small kinetic energies that the $dE/dx$ loss is much larger than for the muons in the beamline; even a very thin window is highly effective in stopping antiprotons (see the discussion on the passage of particles through matter in the PDG, and the rapid rise of the energy loss at small energies shown in the PDG [30]). Annihilations-in-flight in these windows in fact remove far fewer antiprotons than does simple energy loss. Mu2e will use three "absorbers" along the beamline: one at the entrance to the Transport Solenoid, one in the center (as explained in Equation 6), and a third in the middle of the first S-bend.

# 4. THE Mu2e EXTINCTION SYSTEM

## 4.1. Extinction Dipole

Extinction, the suppression of protons outside the beam pulse, proceeds at two levels: beam formation and an external system. Estimates of the extinction from beam formation are about $10^{-(3-4)}$. Mu2e will then use a high-frequency oscillating magnetic field system in the beamline such that the beam passes through when the field is near zero. The field rapidly increases in magnitude so that beam outside the pulse is directed into collimators and does not reach the production target. The "AC dipole" system has 300 kHz and 4.5 MHz AC components (15th harmonic); 50% of the beam is deflected into the collimators at $\pm 87$ ns, with 100% extinction at $\pm 114$ ns. The deflection and the magnetic field over one cycle are shown in **Figure 6**. The transmission, both with and without the AC dipole, is shown in **Figure 6**. The total extinction is better than $10^{-11}$ at all times for beam more than 10 ns outside the nominal transmission window, and the transmission of the in-time beam is estimated to be 99.7%, and the transmission as a function of time is shown in **Figure 7**. Mu2e's requirements are for the extinction to be $10^{-10}$ or better and Mu2e's current post-TDR simulations predict an overall level of $\mathcal{O}(10^{-12})$.

## 4.2. Extinction Monitoring

The calculations of section 4.1 are no substitute for a measurement. Mu2e will measure the extinction in two ways: (1) an upstream monitor before the AC dipole, and (2) a "telescope" downstream of the target. The upstream monitor will be fast and is designed to detect problems in the beam. A thin foil will scatter protons into Cherenkov radiators. Assuming a pre-AC dipole extinction of $10^{-5}$ the system would see tens of particles in about ten seconds and could signal failures on a time scale of minutes. The downstream system looks for protons scattered off the target with a measurement time of order a few hours (assuming a extinction level of $10^{-10}$). The system is depicted in **Figures 8**, **9**. A window in the Production Solenoid allows beam to transit just over the proton beam dump; a permanent dipole selects 4.2 GeV/c protons (value chosen after optimization) that are triggered with scintillators and tracked with pixel planes, with a muon range stack for particle ID following the tracker.

## 4.3. The Mu2e Stopping Target

The Mu2e stopping target consists of about 162 g of > 99.99% pure aluminum. The target location is along the beam magnetic axis as shown in **Figure 3**. The target will consist of 37 foils, each

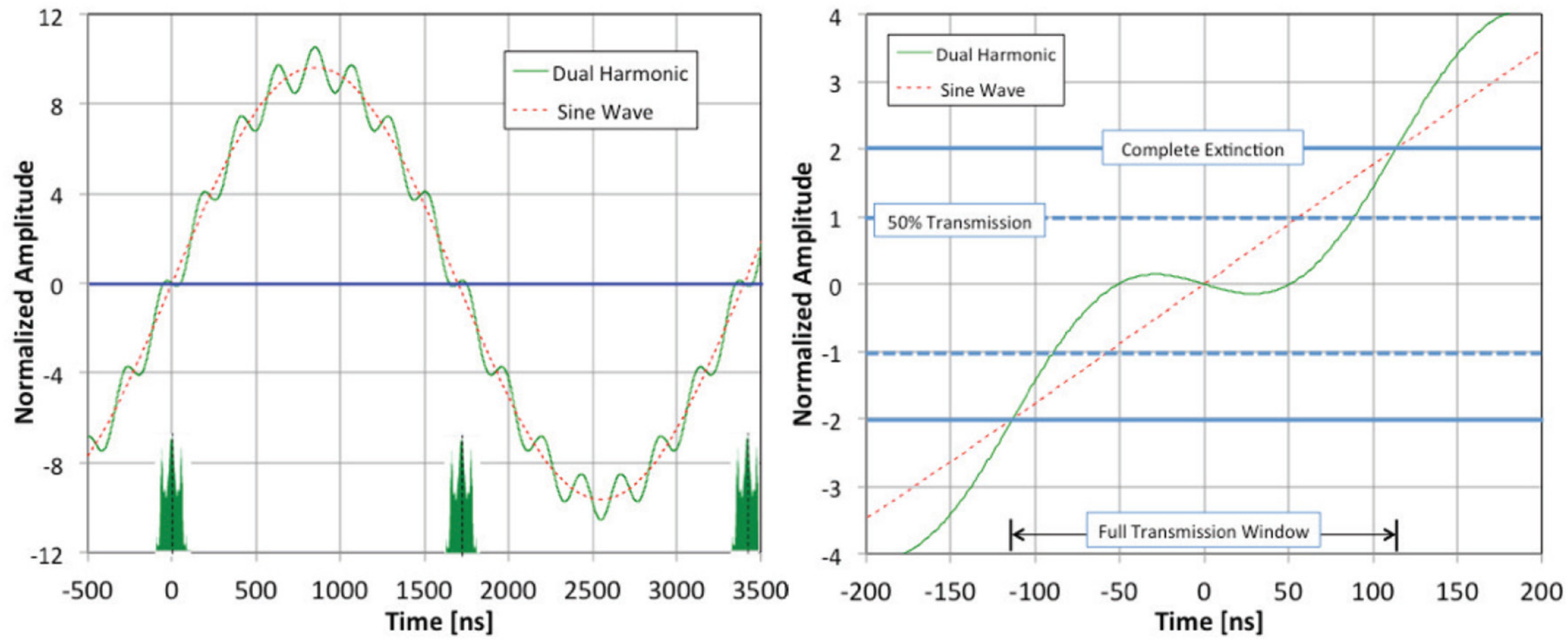

**FIGURE 6 |** Displacement of the beam by the AC dipole vs. time. Recall that the AC dipole system sweeps out-of-time beam into collimators. The left-hand plot shows the normalized amplitude of the field and the right-hand shows the normalized amplitude of the displacement. The units of displacement are "normalized deflection": $\delta = 1$ means the center of the beam is deflected to the edge of the collimators, implying 50% transmission. At $\delta = 2$ the entire beam will be deflected into the collimators. Inset in the left-hand side is the location in time of the proton pulse and its expected shape. Updated from Bartoszek et al. [8], courtesy of E. Prebys and the Mu2e Project.

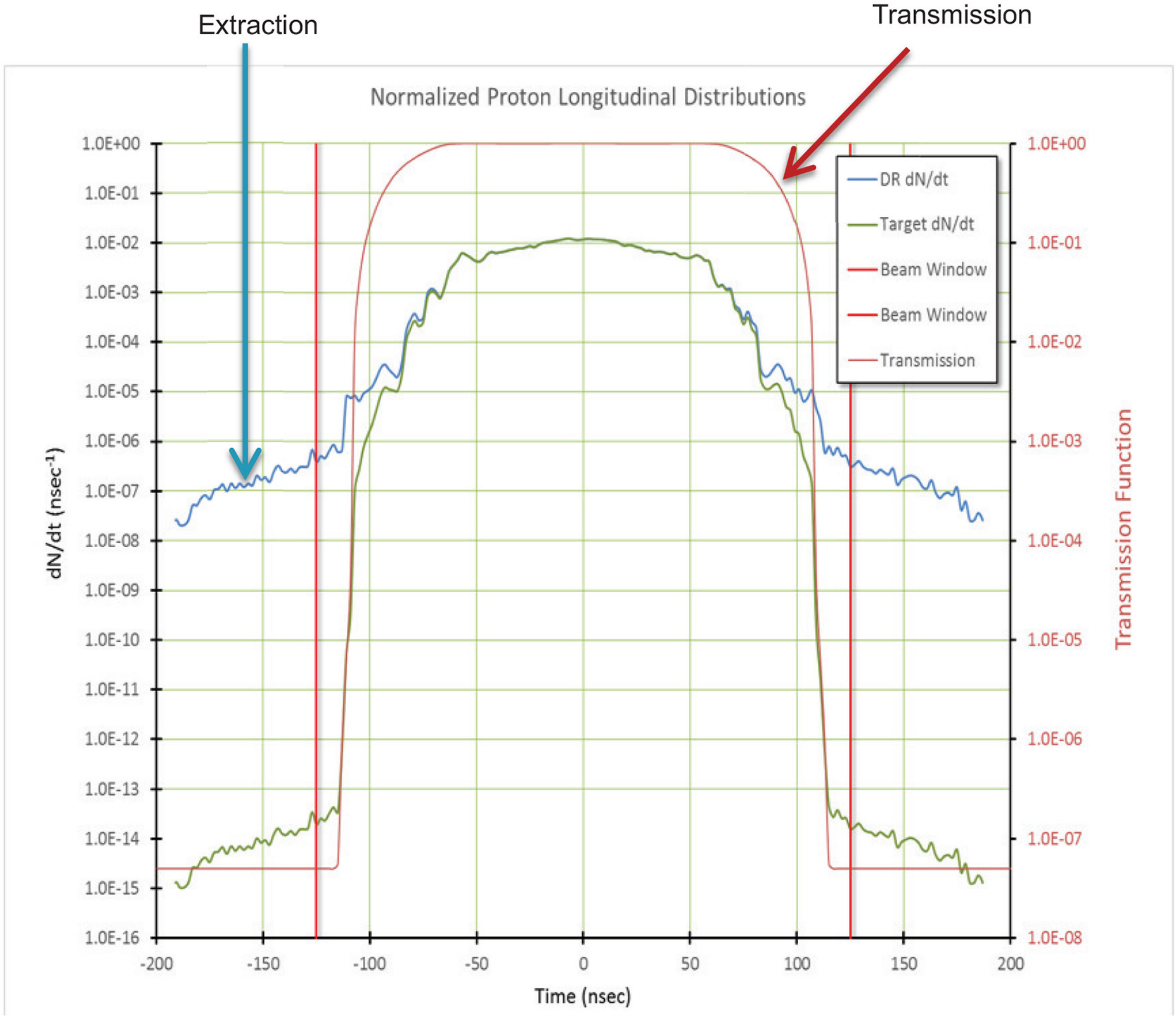

**FIGURE 7 |** Performance of the AC dipole system. The black line (scale at left) shows the transmission curve of the external dipole/collimator system, based on a G4Beamline simulation [31]. The green curve (scale at right) shows the ESME [32] simulation of the beam extracted from the Delivery ring. The blue curve shows the convolution of the two. Updated from Bartoszek et al. [8], courtesy of E. Prebys and the Mu2e Project.

100 $\mu$m thick, with a hole in the center. Recall that the muons are spiraling in the magnetic field so the hole is not a problem; muons passing through the hole of one foil see another as they execute helices. There are three interesting features of the stopping target for our purposes:

- Muons stop in the same material in which they convert, and so outgoing conversion electrons must pass through the stopping material and lose energy by $dE/dx$. This energy loss tends to push electrons down into the DIO region; electrons typically lose an MeV or more. The number and thickness of the foils were therefore carefully optimized. The annular detector has a central hole, and the field is graded; simulations show that the conversion electrons that are most likely to be accepted are produced nearly perpendicular to the beam axis and are pitched forward into the acceptance by the field gradient. If we want to maximize the number of stopped muons while simultaneously minimizing the amount of material seen by outgoing electrons, the best target would be a "gas" uniformly distributed throughout the target volume. At this writing the 37 foil scheme is close to an optimum (although plans involving meshes or flat screens or other arrangements are still being considered).
- As the electron flash passes through the foils, the resultant spray of particles can damage the inner portions of the tracking detector. In the straw tube detector we will discuss in section 5.1 we find charge deposit approaching 1 C/cm for the wires closest to the beam, and have therefore decided on a central hole, making the stopping target annular.
- As muons stop, they eject neutrons and protons that can damage elements of the detector or increase the dead-time of the cosmic ray veto of section 5.3. The stopping target is therefore surrounded by polyethylene absorbers to reduce the flux. Electrons do lose energy and multiple scatter as they pass through this material but it is required to manage the ejected particles.

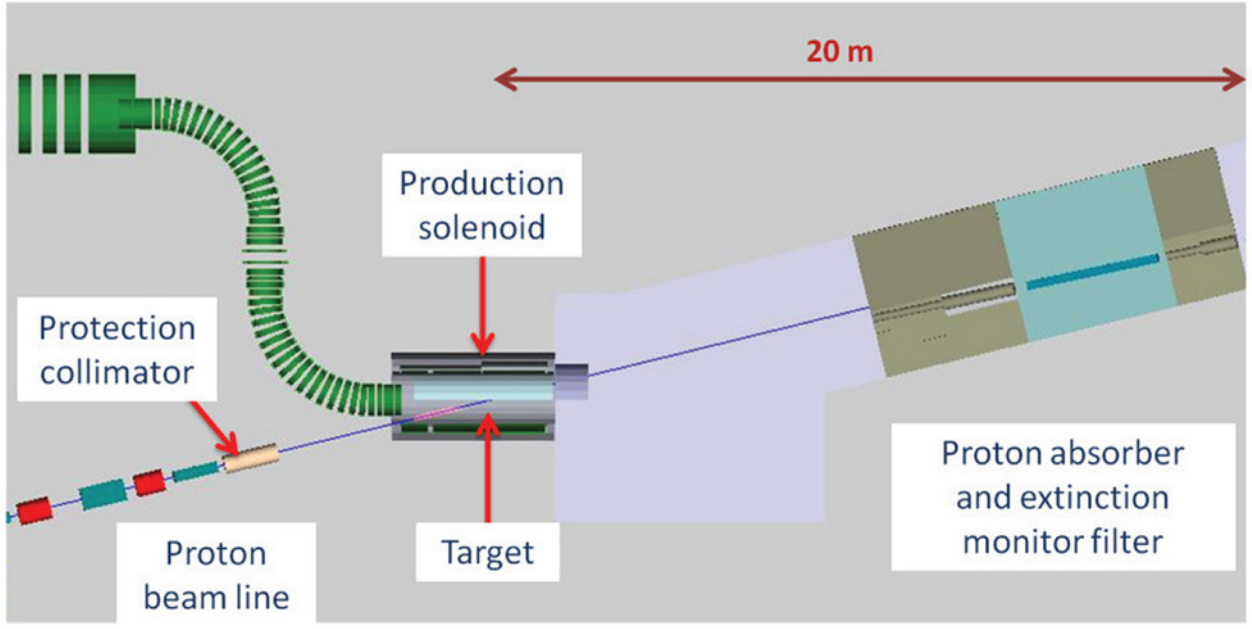

**FIGURE 8 |** Location of the downstream extinction monitor. From Bartoszek et al. [8].

Conversion electrons traveling backwards can be reflected by the increasing gradient and sent back into the tracker. Unfortunately, those electrons have increased energy loss in the stopping target since they pass through it twice; in fact, the contribution of backwards-going electrons to the final accepted sample will be no more than 10–20% (depending on final analysis choices). While this is not negligible, it is far less than the factor of two improvement we would na¨ively expect from the reflection.

## 4.4. The Mu2e Stopping Target Monitor: Measuring the Denominator

Discussions of the experiment rarely touch on the denominator for lack of time; certainly the focus of this experiment will be on detection of a signal, minimization of backgrounds, and the like. Nonetheless, the normalization of the conversion rate to the muon capture rate is an interesting and difficult measurement worthy of explanation.

It is certainly possible to imagine a measurement based on a model of muon production and transmission through the solenoids, but it would be better to base the measurement on a process related to a muon being stopped and captured by a nucleus. The uncertainties from the models of muon production and of the transmission through the solenoid are hard to estimate: a direct measurement is better. Mu2e will use the X-rays emitted as a stopped muon falls into a 1 s muonic state (206 times closer than the electron's Bohr radius: for $^{27}Al_{13}$, about 19 fm compared to the nuclear radius of $\sim$ 4 fm). Muon-to-electron conversion can occur from higher states and can also occur incoherently [33] but both effects are small. The incoherent process is relatively small because the coherent transitions are enhanced as discussed earlier.

Mu2e has chosen to detect the 357 keV X-rays produced as a muon cascades down from the $2p \rightarrow 1s$ state (other lines, such as $3p \rightarrow 1s$ and $4p \rightarrow 1s$ are also observable). The standard choice for detecting these X-rays is a high purity germanium solid-state detector (HPGE) [34]. These detectors are not well matched to the demands of high-energy physics experiments in intense beams. First, HPGEs are too slow to detect all the individual X-rays from muon stops at the nominal intensity, so an event-by-event count is not possible. Second, the Mu2e beam has an intense electron component from the beam flash. Mu2e has two problems associated with the flash: (1) the same aluminum that stops muons will then produce bremsstrahlung photons at a rate we calculate to be 51 MHz/cm$^2$ with a mean energy of 1.4 MeV and (2) many of these photons are above pair production threshold, so if they strike the HPGE detector they can produce $e^+e^-$ pairs that can cause radiation damage. The calculation of the damage is complicated and uncertain, since most damage calculations are done for neutrons—and then one needs to understand the physics of non-ionizing energy loss (NIEL) and how to correctly add up the different sorts of damage. The interesting muons arrive about 100 ns after the flash and produce their X-rays within picoseconds of their arrival. Commercial off-the-shelf detectors and their electronics cannot manage these rates (MeV/sec limits) and their associated electronics are not fast enough. The radiation damage to a detector placed just after the Detector Solenoid would require annealing, usually after around $10^9$ n/cm$^2$, in a matter of hours to days unless the experiment takes measures to prevent this. The annealing cycle requires heating the detector to remove lattice imperfections (as in the discussion of section 3.1.1) and takes more than a day to complete, making this short time impractical. The detector needs to be about 35 m from the stopping target (reducing the rate by $\sim 1/r^2$, $r$ from the stopping target to the detector) and heavily shielded in order to make the radiation damage and rates manageable. The final design is under active development at this writing. An alternative solution is to detect delayed photons from $^{27}Al \rightarrow {}^{27}Mg$ (13% of muon captures). The excited $^{27}Mg$ beta-decays to an excited state of $^{27}Al$ with a half-life of 9.5 min. The experiment could use the macrostructure of the beam to wait until a long "off" period (0.5 s on, 0.8 s off) thereby avoiding rate problems. The excited $^{27}Al$ quickly (ps) emits an 844 keV photon that can then be detected. This method, like the direct detection of the $2p \rightarrow 1s$ X-rays, cannot provide an event-by-event measurement: even if one could overcome the shielding and flash problems, the rate is only 13% and the technique averages over the 9.5 min half-life.

## 5. THE Mu2e DETECTOR

Mu2e's detectors are annular: the detectors are in a solenoidal field of about 1 T along the $z$-axis and the muon beam is along this direction. There are two reasons for this design:

- The experiment will be exposed to $\mathcal{O}(10^{10})$ $\mu$/sec and $\mathcal{O}(10^{18})$ $\mu$ over the life of the experiment. The products of muon capture, remnant beam, and electrons produced from the initial proton collision would overwhelm any detector through both instantaneous occupancy and accumulated radiation damage. The annular design allows the passage of these products to the beam dump without striking detector elements.
- The decays of muons, occurring more evenly throughout the cycle, because of their longer lifetime, are typically too low momentum to exit the central region: $p_\perp \propto qBR$ in a solenoidal field. We saw the electron spectrum in section 2.2 and only $\mathcal{O}(10^{-12})$ of the decays have $R$ large enough to be reconstructed. Most muons never reach a detector element, reducing the number of detector hits and found tracks to a manageable level.

COMET has chosen to put a bend in their detector solenoid system to serve the same purposes [6]. Each of these choices has advantages and disadvantages. The backgrounds from RPC and RMC arise from converted photons. Mu2e, by measuring the number and spectrum of observed $e^+$ the experiment can, *in situ*, estimate the number of $e^-$. The charge-symmetry also permits Mu2e to simultaneously search for $\mu^- N \rightarrow e^+ N'$ [13, 14]. COMET Stage II, with a momentum-selecting bend, does not see $e^-$ and $e^+$ simultaneously but does not need a hole, reducing the problems from radiation damage and accidental activity in the detector.

The detector consists of the standard arrangement of a tracker followed by a calorimeter, surrounded by a cosmic ray veto. A schematic of the detector region is given in **Figure 10**.

## 5.1. Tracker

The Mu2e tracker is the most important detector element. The tracker must have as good resolution as is reasonably achievable in order to minimize backgrounds, especially from the decay-in-orbit electrons. Since the DIO spectrum falls as $(E_{\text{conversion}} - E)^5$ near the endpoint, the background increases quickly as the resolution degrades. Further, since the DIO spectrum is *below* the endpoint, high-side tails in the resolution function are particularly problematic. The experiment also wants to stop as many muons as possible in as little material as possible, since $dE/dx$ in the stopping target smears the monochromatic energy distribution of the conversion electron downwards. It is true that DIO events near the conversion peak are smeared down as well, but this is a stochastic process and conversion electrons with relatively large energy loss can end up in the portion of the DIO spectrum that underwent relatively small energy loss. This effect implies the energy loss in the tracker needs to be as small as possible (which is why, as noted earlier, the stopping target only stops about 40% of the muons). This combination of a requirement for high core resolution, small high-side tails on the resolution, and minimum energy loss has led the experiment to choose a straw tube tracker.

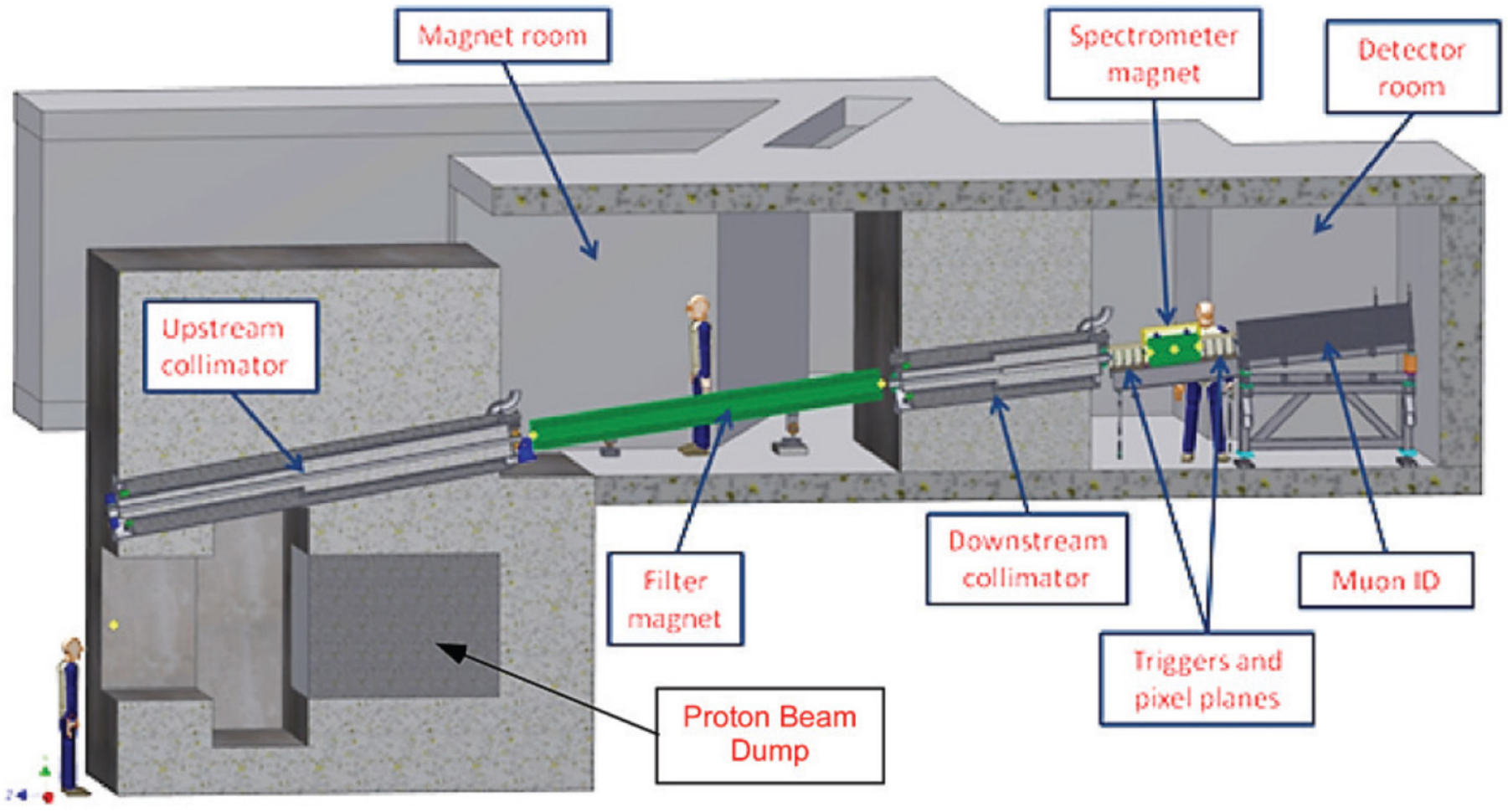

**FIGURE 9 |** Schematic of the downstream extinction monitor. From Bartoszek et al. [8].

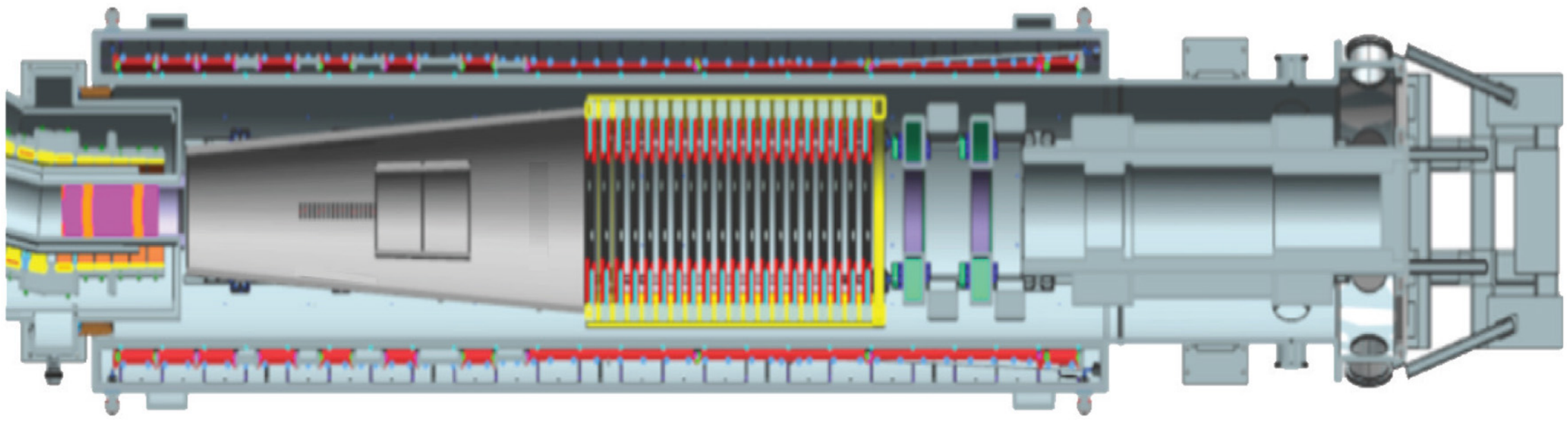

**FIGURE 10 |** Cutaway view of detector. The stopping target and its surrounding polyethylene absorbers are on the left. The tracker follows, with the two disks of the calorimeter to the right. A muon beam stop is indicated after the calorimeter, and the solenoid endcap is to the right. The detector elements and beam stop are placed on rails so that they can be assembled externally to, or removed from, the detector solenoid. Figure courtesy of the Mu2e collaboration.

Straw tubes offer an excellent combination of low mass, short drift times, and excellent resolution and are in wide use in particle physics (see the PDG, [30], for an overview). Still, the Mu2e requirements are exceptionally stringent.

The overall geometry of the tracker system is a series of 18 stations along the beam axis, shown in **Figure 11**. The design here is the one presented in Lucà [37], slightly advanced from the design in Bartoszek et al. [8] and the details of the design continue to evolve slightly. The drift gas will be 80:20 Argon:$CO_2$ at an operating voltage of 1,500 V. The basic tracker element is a 25 $\mu$m gold plated tungsten sense wire centered in a 5 mm diameter tube, referred to as a straw. Each straw is made of two layers of 6 $\mu$m (25 gauge) Mylar®, spiral wound, with a 3 $\mu$m layer of adhesive between layers. The total thickness of the straw wall is 15 $\mu$m. The inner surface has 500 Å of aluminum overlaid with 200 Å of gold as the cathode layer. The outer surface has 500 Å of aluminum to act as additional electrostatic shielding and to reduce the leak rate. The straws vary in active length from 334 to 1,174 mm and are supported only at the ends. Groups of 96 straws are assembled into panels. Each panel covers a 120° arc and has two layers of straws to improve efficiency and help determine on which side of the sense wire a track passes (the classic left-right ambiguity: one measures the time that a wire is struck relative to some other time, providing a distance but not a direction). A 1 mm gap is maintained between straws to allow for manufacturing tolerances and expansion due to gas pressure. This necessitates that individual straws be self-supporting across their span. The tracker consists of 18 stations, evenly spaced along its whole length of 3 m, and associated infrastructure. Each station is made of two planes (36 planes total) and a plane consists of 6 panels (216 panels total) rotated by 30°, on two faces of a support ring;

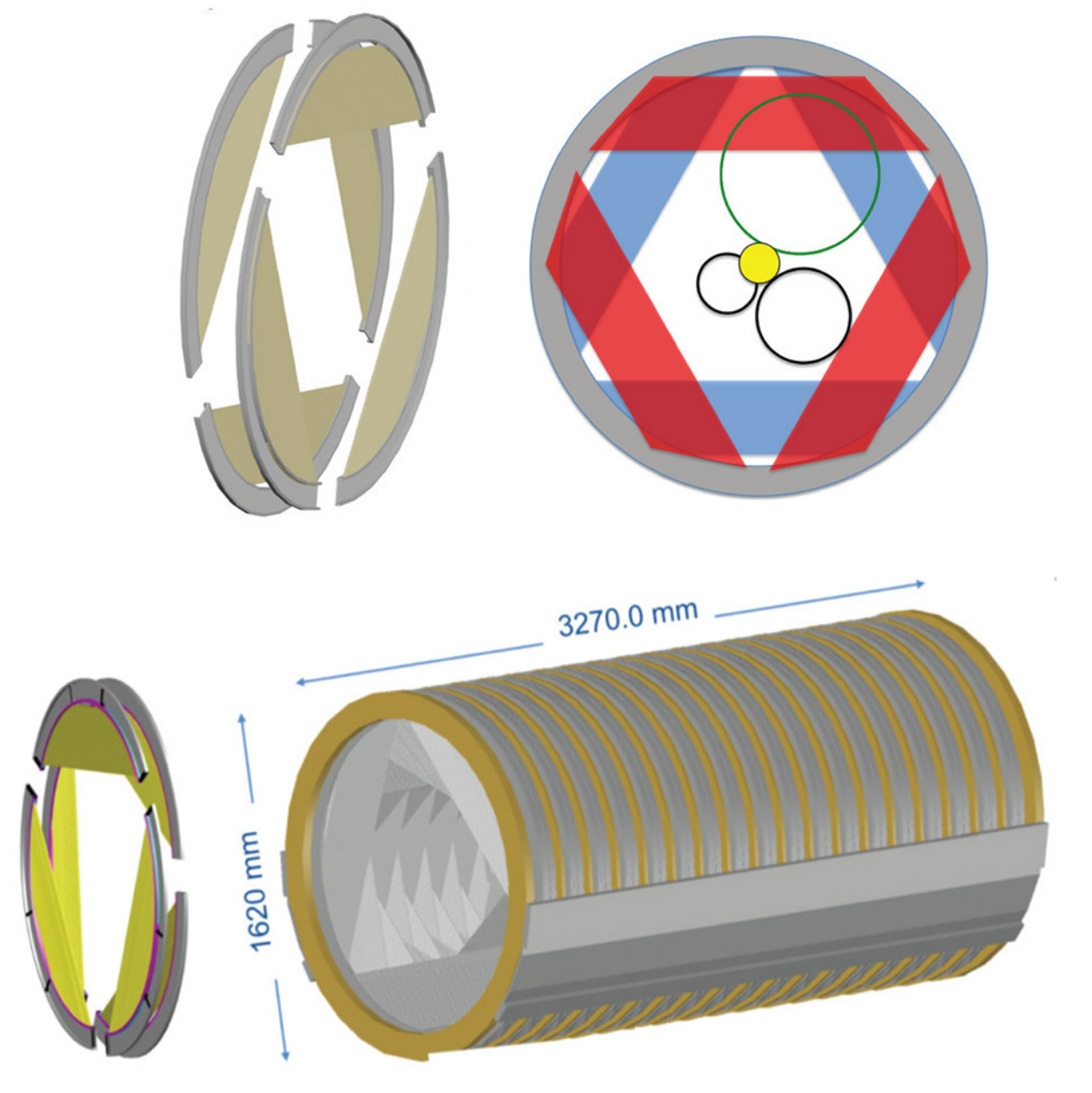

**FIGURE 11 |** The Mu2e tracker. The upper left picture shows panels assembled into a plane and a station. The assembled tracker is shown in the bottom figure. The upper right shows a beam's-eye view of a station: the three circles are projections of tracks at the Michel peak (small black circle), an intermediate momentum, and the conversion energy (last in green). Figures from Miyashita [35].

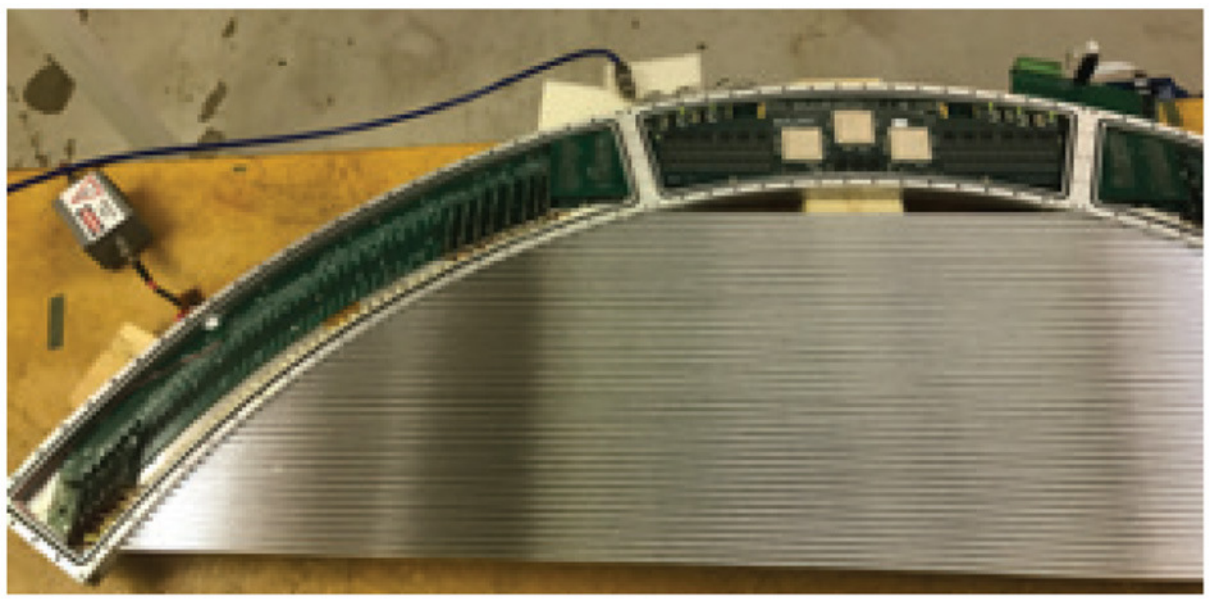

**FIGURE 12 |** One panel and the front end electronics for the Mu2e tracker. Figure from Bonventre [36].

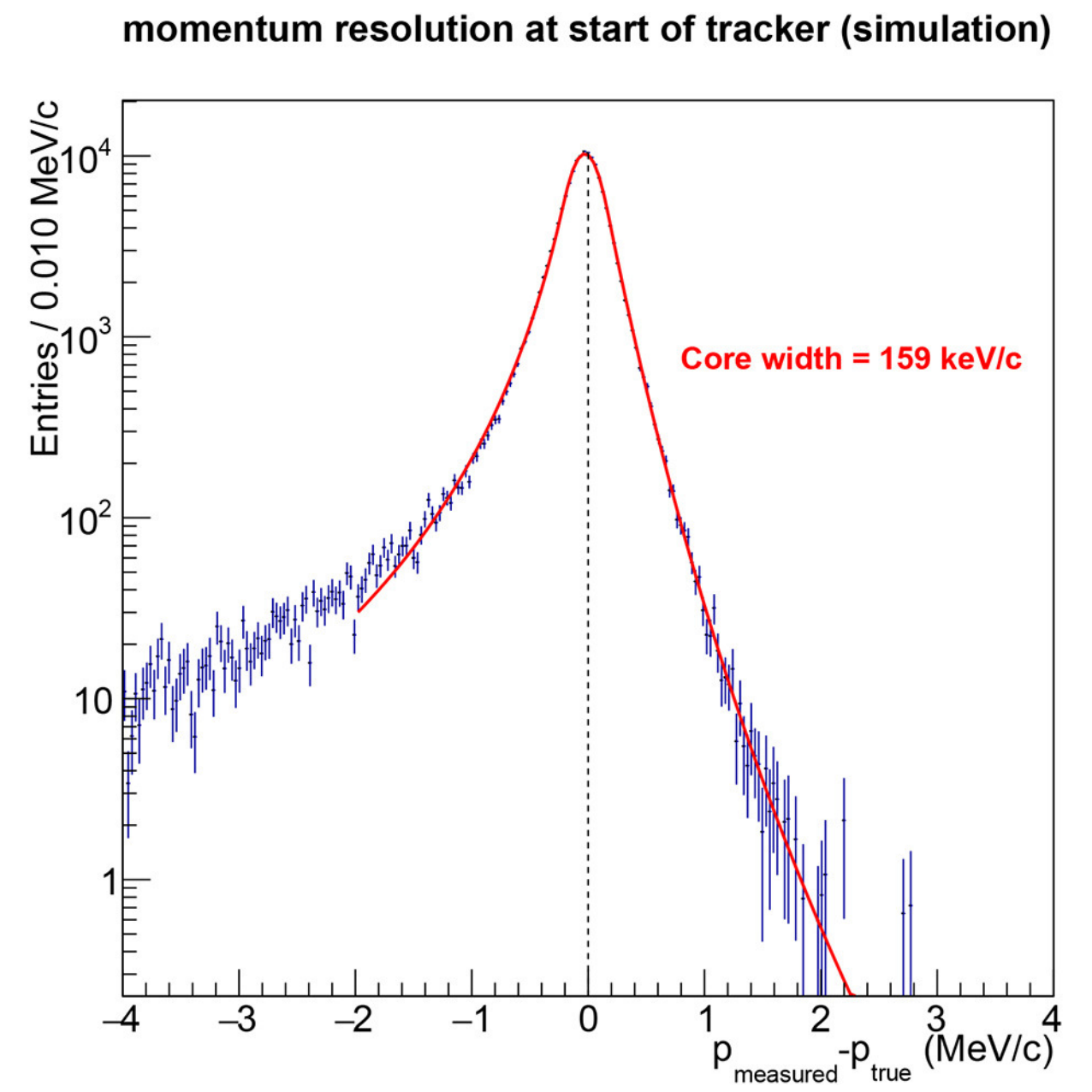

**FIGURE 13 |** The resolution of the Mu2e tracker for electrons at the conversion energy. The sample is chosen with cuts approximating the final data sample. The asymmetric low side tail is because of the stochastic nature of energy loss in the tracker. Note that the high-side tail, where decay-in-orbit events would be "promoted" to the signal region, is small. The simulation uses measured properties of the tracker, charge cluster formation, processing by the electronics, and passage through simulated DAQ, which is then passed to a reconstruction program that includes accidental activity from the beam and stopped muons. "Core resolution" refers to a fit to the central part of the resolution and can be thought of as the Gaussian $\sigma$. Courtesy of the Mu2e collaboration.

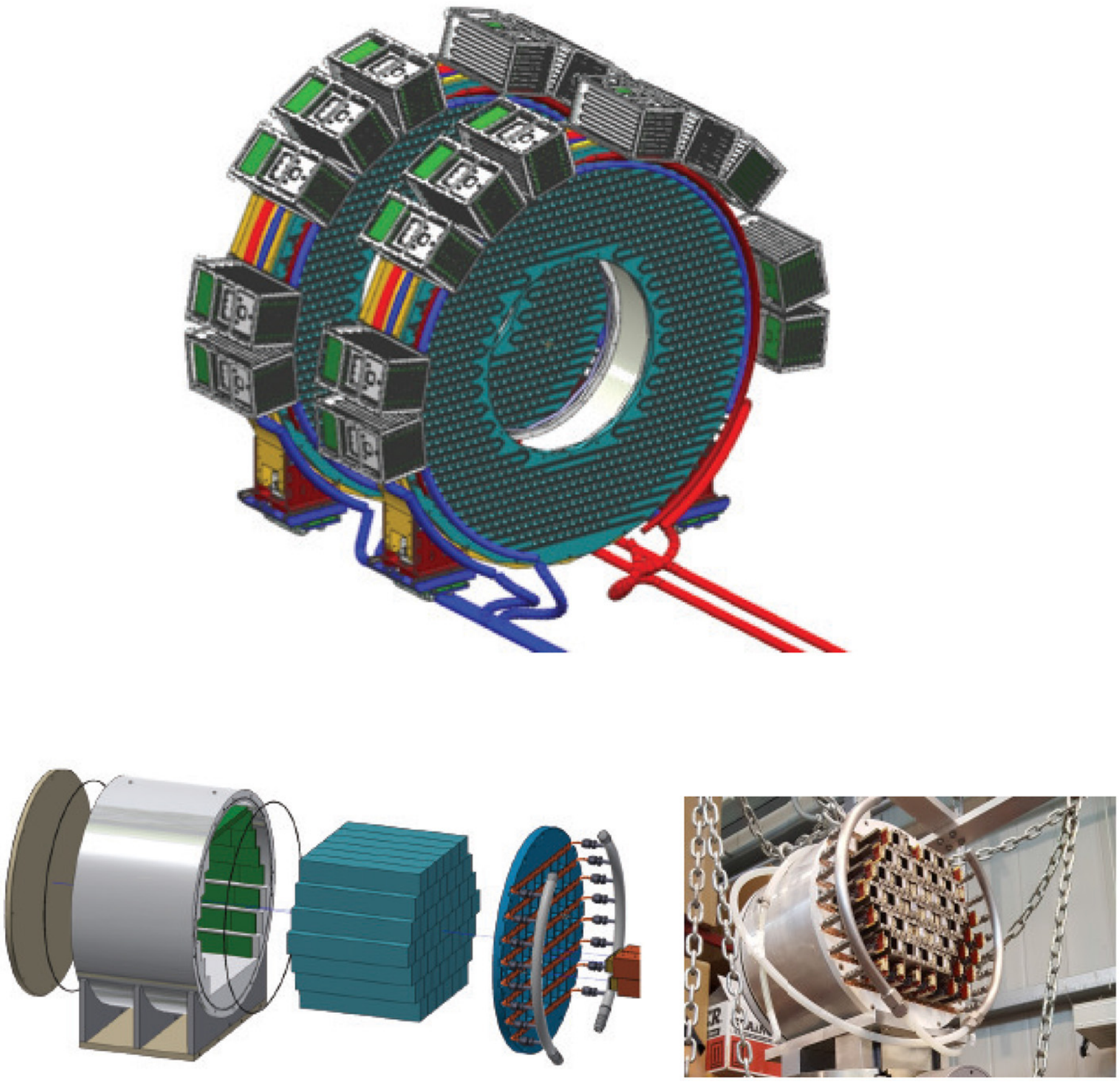

**FIGURE 14 |** A schematic of the Mu2e calorimeter showing the two disks, location of readout modules, and part of the calibration system. The bottom half of the Figure shows the assembly of the test module and a picture of the test module being moved into place.

there are three panels per face. The tracker has a total of 20,736 straws.

Groups of 96 straws are assembled into panels. One panel is shown in **Figure 12**. The straws are visible, as are the front end electronics. Note the electronics is on the outside of the tracker; radiation-hard FPGAs will be used. Each panel covers a 120° arc and has two layers of straws to improve efficiency and help determine on which side of the sense wire a track passes (the classic left-right ambiguity). The straws use two-sided readout and a comparison of the arrival time at the two ends of the straws to determine the position along the straw; the resolution here is about 4 cm, more than sufficient for Mu2e's purposes given that a track, spiraling through the detector in the solenoidal field, hits many straws.

As we have stressed, the tracker is annular, with a central hole for passage of the muon beam. Electrons from Michel decays of free muons have a maximum momentum of 52.8 MeV/c and their radius in the $\approx$ 1 T magnetic field is too small to produce hits (as was shown in **Figure 11**). Most of the higher momentum decay-in-orbit electrons also have too low a momentum to be successfully reconstructed as well; only a few hundred thousand are seen, making a measurement of $R_{\mu e}$ at the $10^{-17}$ level possible since then one rejects only $\mathcal{O}(1/10^5)$ not $\mathcal{O}(1/10^{17})$. The expected resolution of the tracker is shown in **Figure 13**.

## 5.2. Calorimeter

The Mu2e calorimeter serves several purposes: (1) particle identification, specifically $e/\mu$ separation to remove muons with the electron signal momentum; (2) improving the tracker reconstruction, by providing a "seed" for reconstruction as well as a consistency check; (3) a standalone trigger for the experiment. The calorimeter consists of two disks, with a central hole for passage of the remnant muon beam and the beam flash. The separation between the two disks is specifically chosen to be "half a wavelength" for the 105 MeV/c conversion electron in the 1 T field: if a conversion electron passes through the hole at the center of the first disk, it will hit the second. A schematic of the calorimeter is shown in **Figure 14**; also shown is a test module that provides data we will present later.

The calorimeter needs

- an energy resolution $\sigma_E/E < 10\%$
- timing resolution $\sigma_t < 500$ ps
- position resolution $<$ 1 cm
- to work in a vacuum of $10^{-4}$ Torr
- and a 1 T Magnetic Field

Each calorimeter disk will have 674 undoped CsI crystals, 34 $\times$ 34 $\times$ 200 mm$^3$ and will be read out with two UV-extended

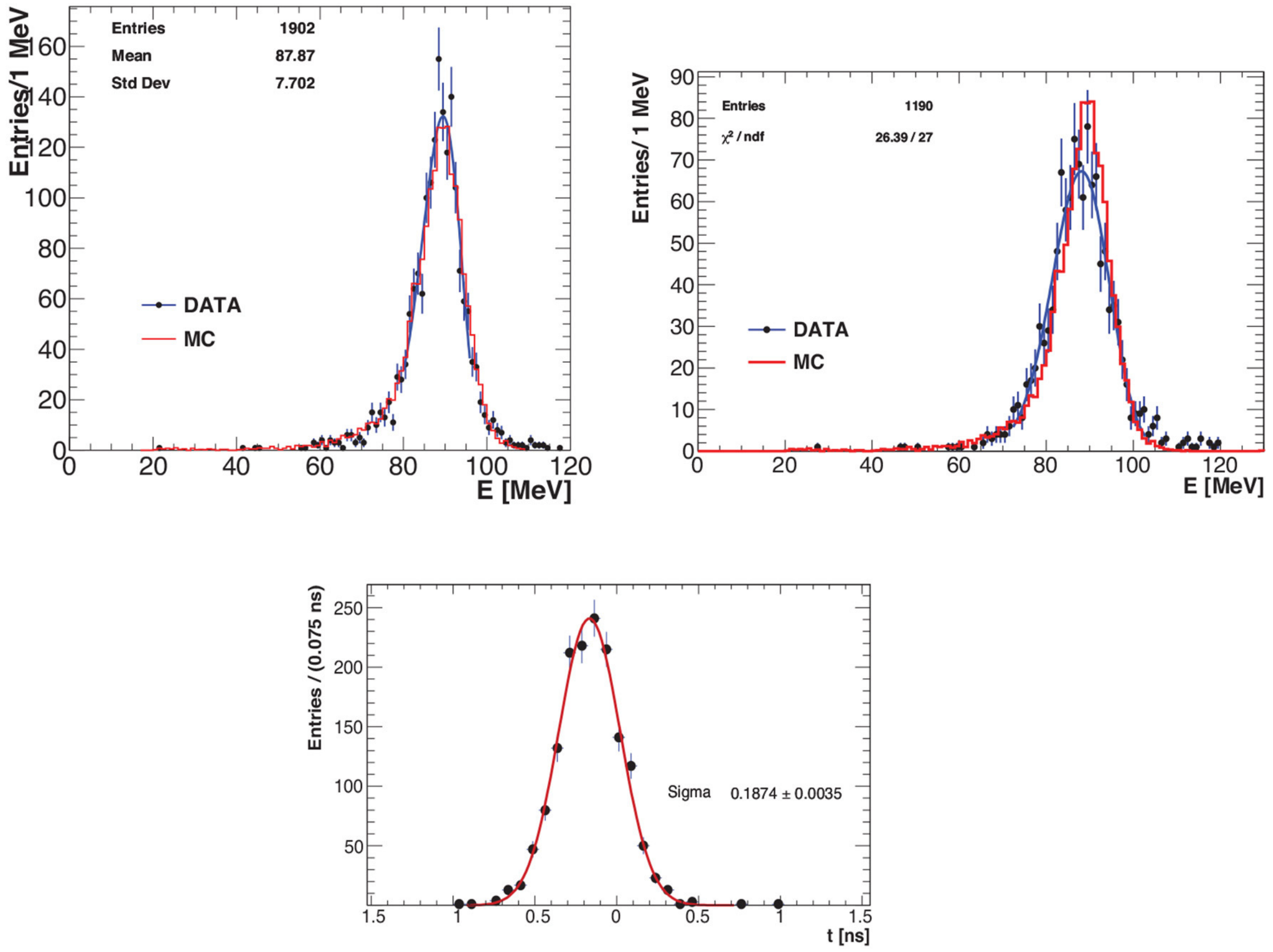

**FIGURE 15 |** The energy and time resolution of the test module for the Mu2e calorimeter for electrons at the conversion energy. The top left is for electrons striking the array at normal incidence; the top right, for 55° to the face. A simulation, using measured properties to tune GEANT4, is shown, indicating good agreement.

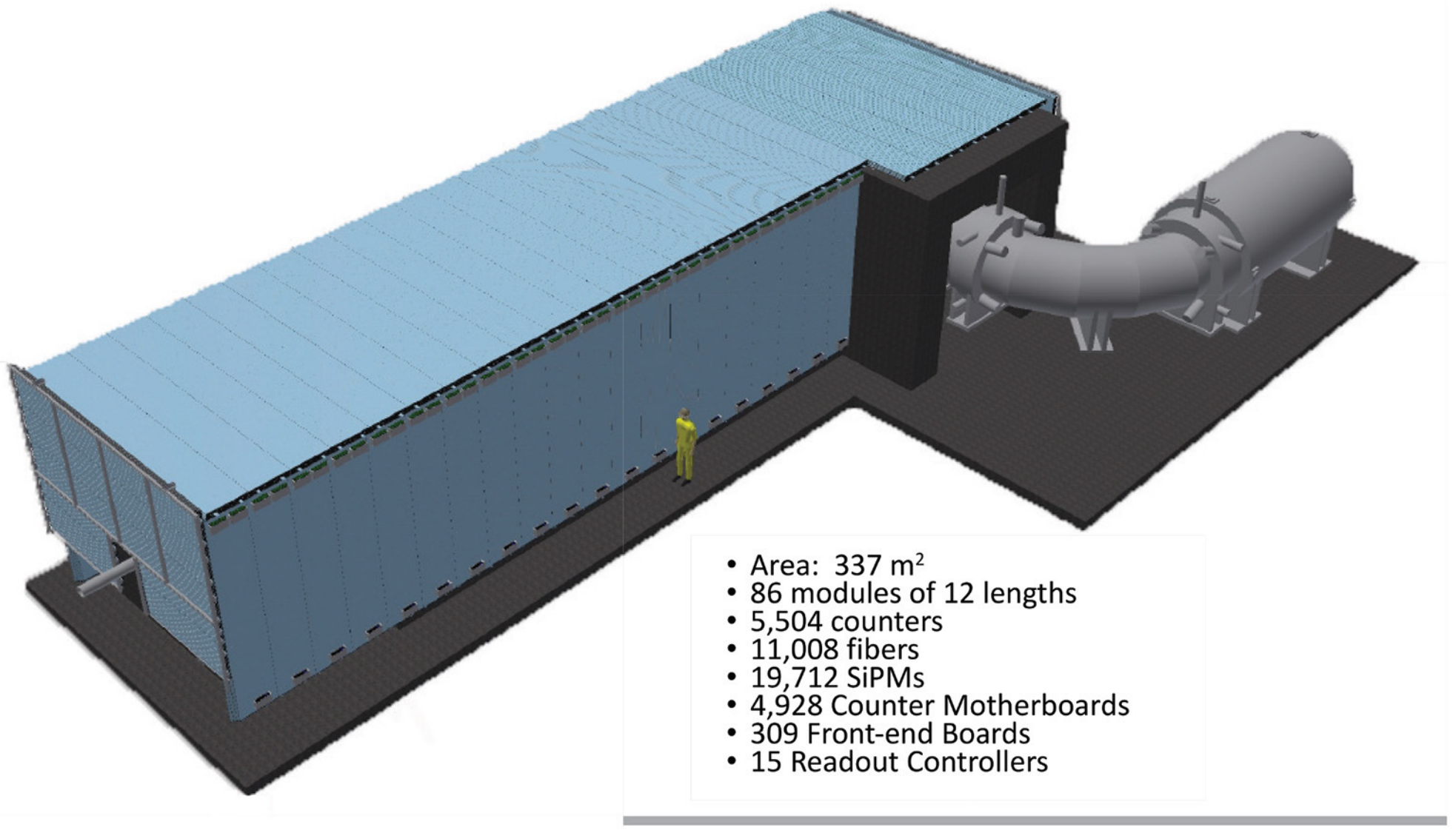

**FIGURE 16 |** The Mu2e CRV system, showing the coverage and scale. The Production Solenoid is to the right; the system covers the Detector Solenoid and part of the Transport Solenoid. From Bartoszek et al. [8].

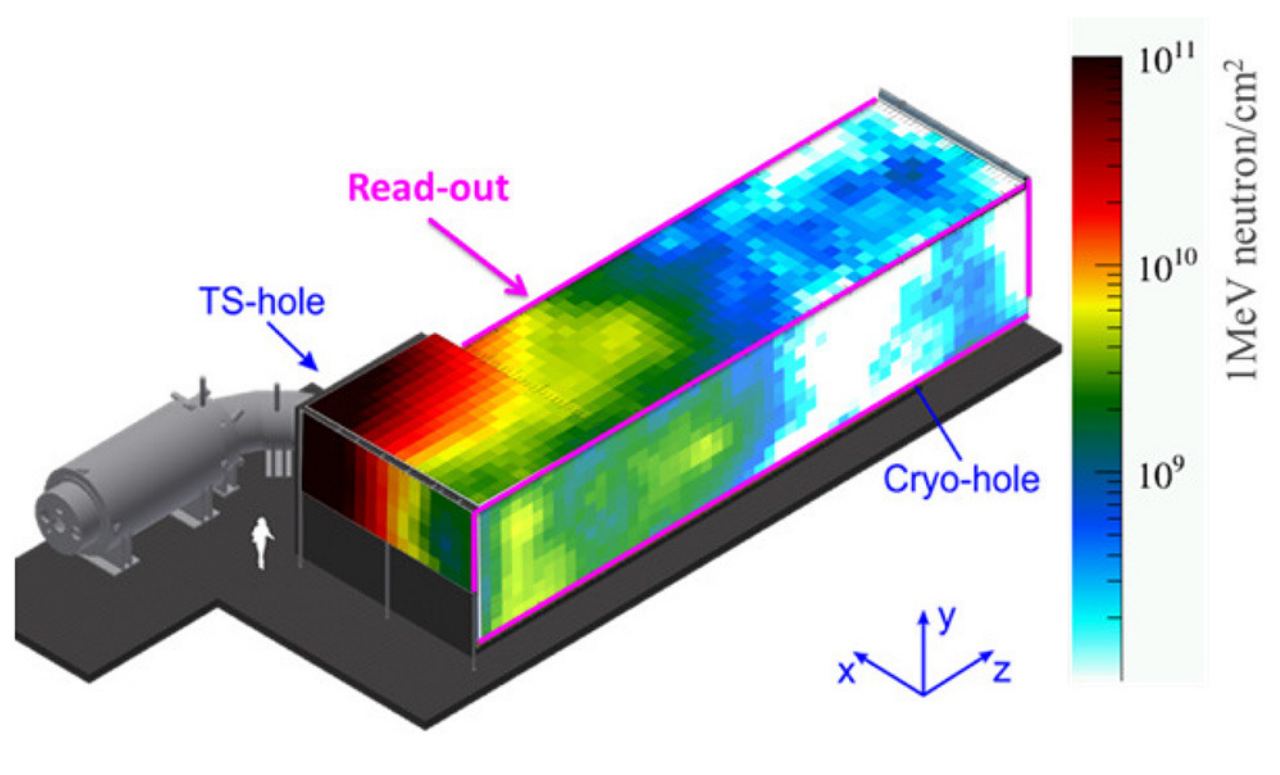

**FIGURE 17 |** The Mu2e CRV system, showing the neutron rates over the counters. From Bartoszek et al. [8].

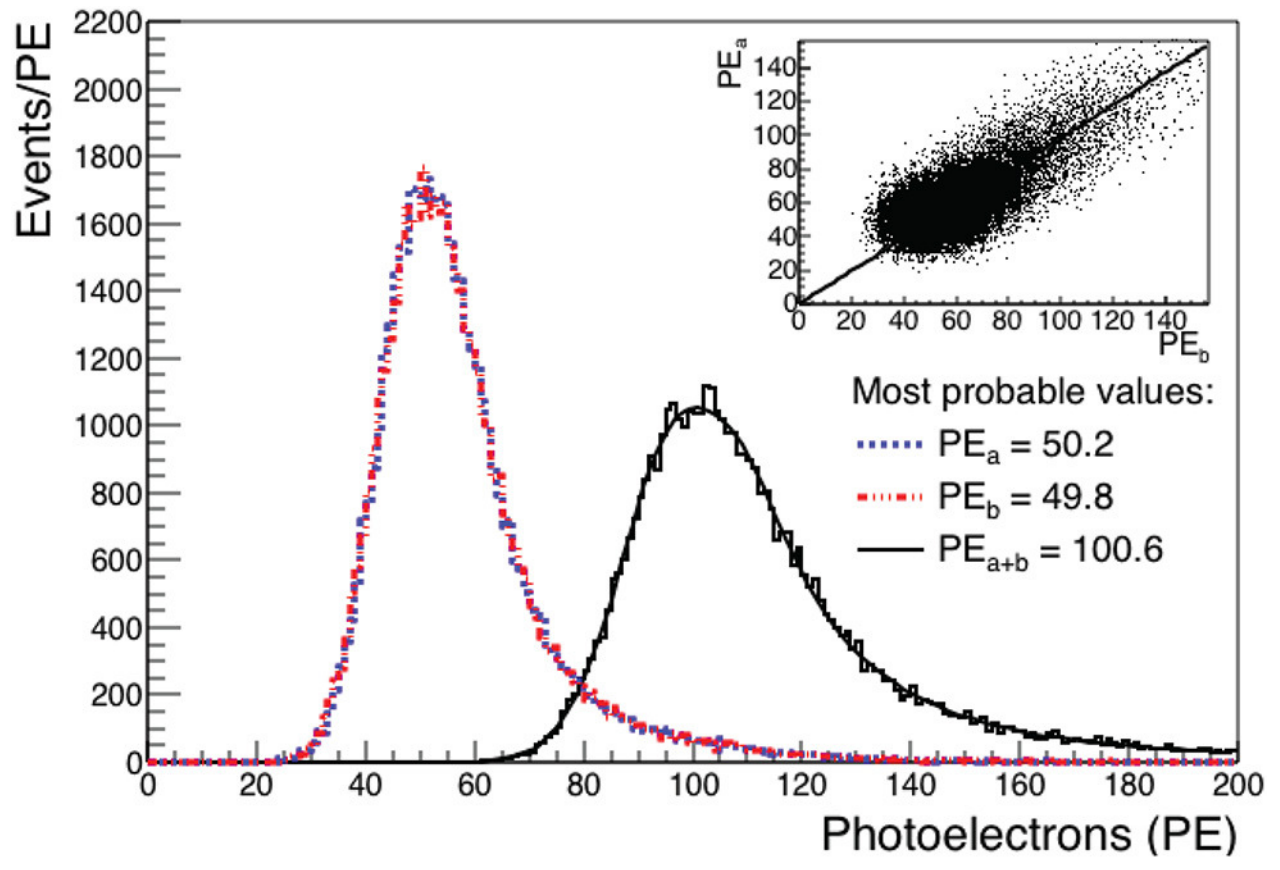

**FIGURE 18 |** Response in photoelectrons at a position 1 m from the SiPM readout for the reference counter read out with 2 × 2 mm$^2$ SiPMs. Dashed and dotted curves are the respective responses from each of the two SiPMs at one end of the counter, and the solid line is the sum of the responses. The distribution was fit to the sum of a Gaussian and a Landau function and the fit is also shown for the summed response. The inset shows the correlation between the two channels and the line from the correlation fit described in Artikov et al. [39]. Figure and caption text from Artikov et al. [39].

Hamamatsu SiPMs. A 51-crystal prototype module was exposed to a test beam and results for energy and time resolution are shown in **Figure 15**, discussed in Atanov et al. [38]. One sees good agreement between data and the simulation for two incident angles; because of the solenoidal field, incident electrons enter the calorimeter at about 55°. The reason for the low-side tail is simply shower leakage, both out the back and "splashback" at the front face. A 20 cm CsI array is only 10.75 $X_o$, so the leakage is unavoidable, but the length is sufficient for the needs of the experiment.

## 5.3. Cosmic Ray Veto

Like many rare process experiments, Mu2e needs a cosmic ray veto (CRV) system. In Mu2e, cosmic rays can produce signal-like events in two ways:

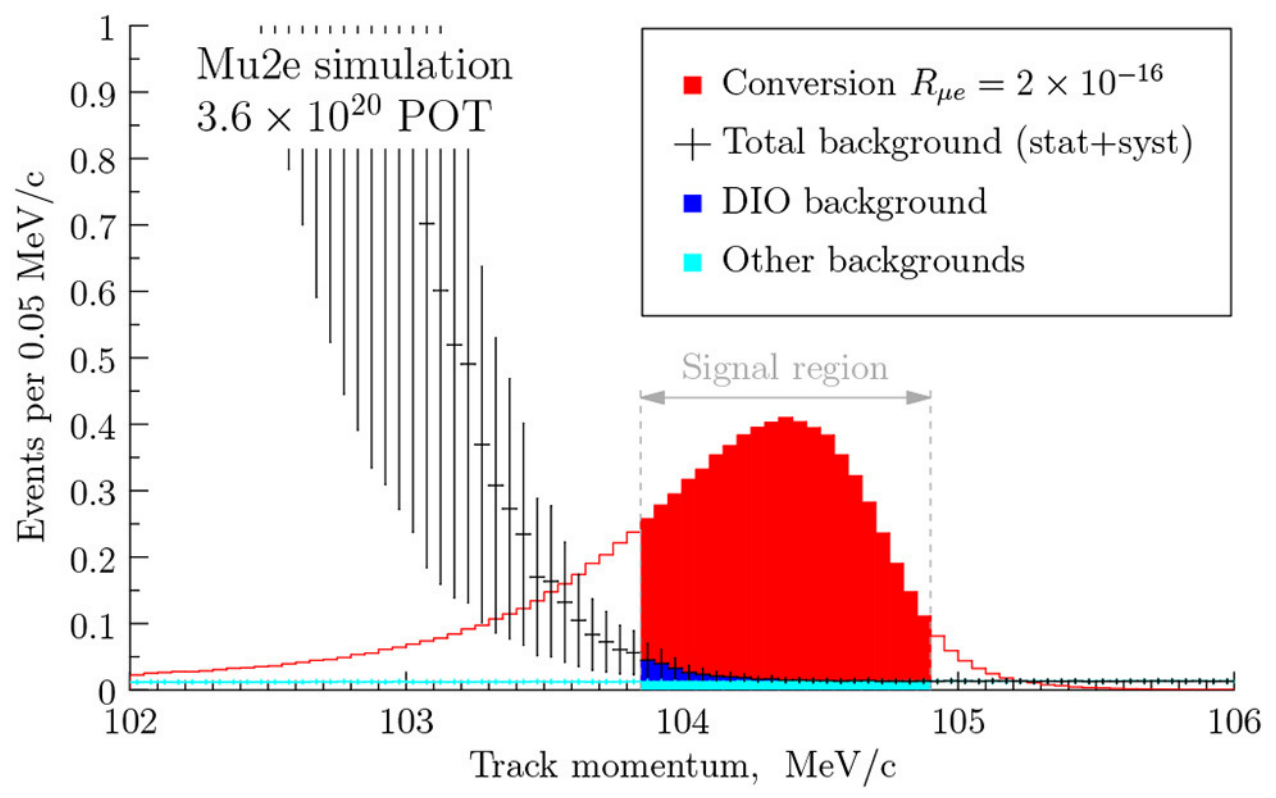

**FIGURE 19 |** Signal and backgrounds for nominal Mu2e exposure, for a signal of $R_{\mu e} = 2 \times 10^{-16}$. The shape of the conversion signal, and the reason its mean is below the conversion energy, is largely due to $dE/dx$ in the stopping target, as explained in section 4.3.

**TABLE 1 |** Backgrounds in Mu2e for the nominal 3.6 × 10$^{20}$ protons-on-target.

| Process | Expected number |
|---|---|
| Cosmic ray Muons | $0.209 \pm 0.02 \pm 0.06$ |
| DIO | $0.144 \pm 0.03 \pm 0.11$ |
| Antiprotons | $0.040 \pm 0.001 \pm 0.020$ |
| RPC | $0.021 \pm 0.001 \pm 0.002$ |
| Muon DIF | $< 0.003$ |
| Pion DIF | $0.001 \pm < 0.001$ |
| Beam electrons | $2.1 \pm 1.0 \times 10^{-4}$ |
| RMC | $0.000^{+0.004}_{-0.000}$ |
| Total | $0.41 \pm 0.03$ |

- A cosmic ray muon striking the stopping target can knock out an electron in the signal region. Since the electron comes from the stopping target and heads down the same path as a signal electron, there is no way to reject the event from measurements in the detector. Perhaps surprisingly, this source would result in $\mathcal{O}(1)$ event/day, or $\mathcal{O}(1{,}000)$ over the lifetime of the experiment.
- Cosmic rays can also result in electrons with approximately the conversion energy knocked out further upstream in the beamline; if the electron is then trapped in the field it can propagate through the stopping target region and be reconstructed. Again, there is no way to separate these from signal electrons.
- Cosmic ray muons can decay-in-flight into electrons while in the solenoids.
- Cosmic ray muons can be misidentified as electrons (although comparing the tracker to the calorimeter information can suppress these with particle identification).

Mu2e thus requires a cosmic ray veto system that covers not just the detector region, but well back into the Transport Solenoid. This upstream region is problematic because of the neutron flux born in the original 8 kW production target. Radiation damage to the detector and readout is an issue. The experiment chose extruded scintillator with embedded wavelength-shifting fibers since the technology is relatively cheap, robust, and uncomplicated, requiring little maintenance. This neutron flux can also produce a large deadtime from interactions in the scintillator—Mu2e considered resistive plate chambers, which are relatively "neutron blind" but the required gas system was inaccessible and not easily repaired since it would be embedded in shielding. Therefore, in addition to the considerable shielding problems of the Production Solenoid, the design of shielding for the CRV itself is a complicated design problem: a system that is 100% dead is not of much use.

The reader is referred to Dukes and Ehrlich [40] for details; only an overview is given here. There are 5504 counters as depicted in **Figure 16** over 327 $m^2$. The counters have embedded wavelength-shifting fibers; each fiber is read out on both ends by a $2 \times 2$ $mm^2$ SiPM. The veto inefficiency has to be 0.01% or less, and the experiment will use four layers of scintillators, separated by Al plates, with layers offset to minimize the effect of gaps. With a 3/4 requirement, the single-layer inefficiency is then $<$ 0.5%. **Figure 17** shows the rates from the neutron flux described above. The average SiPM hit rate is 44 kHz during the live gate. Neutron rates, shown in the figure, are within limits. The deadtime is estimated to be $\approx$ 5%. The measured light yield is shown in **Figure 18** to be 42 photo-electrons for a normally incident muon [39].

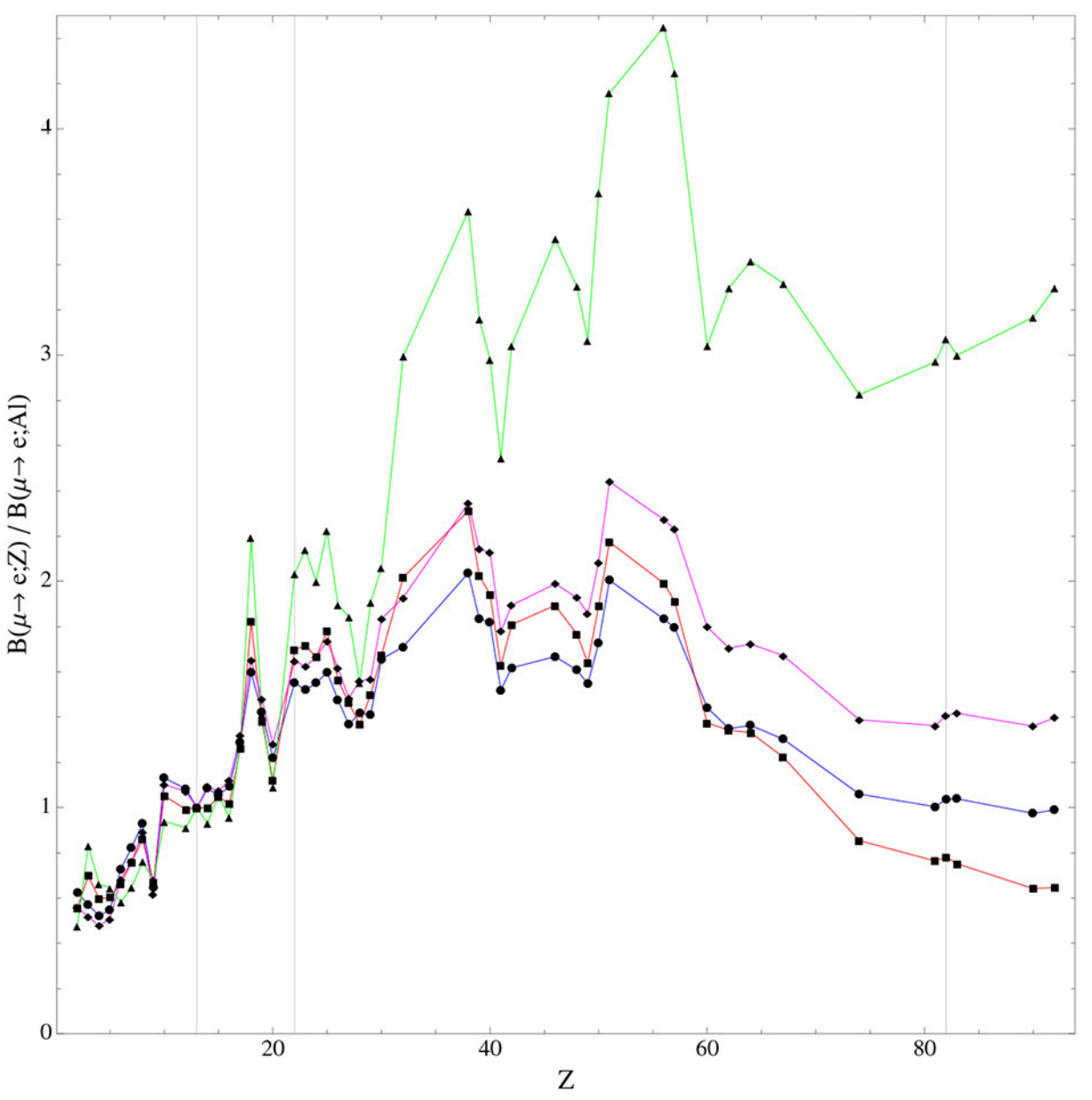

**FIGURE 20 |** Figure and caption from Cirigliano et al. [44]. Target dependence of the $\mu \to e$ conversion rate in different single-operator dominance models. We plot the conversion rates normalized to the rate in Aluminum ($Z = 13$) vs. the atomic number $Z$ for the four theoretical models described in the text: $D$ (blue), $S$ (red), $V^{(\gamma)}$ (magenta), $V^{(Z)}$ (green). The vertical lines correspond to $Z = 13$ (Al), $Z = 22$ (Ti), and $Z = 83$ (Pb).

## 6. EXPECTED SENSITIVITY

Mu2e plans to accumulate $3.6 \times 10^{20}$ protons-on-target. The signal and backgrounds for a signal at the $5\sigma$ discovery level (determined by the method of Feldman and Cousins [41]) as of this writing is shown in **Figure 19**. **Table 1** presents a table of backgrounds. A $5\sigma$ discovery requires $\approx$ 7.5 events against the estimated background of 0.41.

## 7. Mu2e UPGRADES

Abusalma et al. [27] considers upgrades to Mu2e. Advancing the Mu2e technique by an order-of-magnitude is a compelling goal: if Mu2e sets a limit, it will exclude a wide amount of parameter space for many models, and a $\times 10$ improvement would place significantly greater constraints; in the case of an observed signal, such an experiment could explore the nature of new physics. Fermilab's PIP-II could provide an 800 MeV beam with an appropriate time structure; an advantage of 800 MeV protons is that they are far below the threshold for antiproton production and that uncertain background disappears. Neuffer [42] is a sample beam delivery scheme and the PIP-II project is summarized in Ball et al. [43].

One natural upgrade is to replace the aluminum target by titanium. **Figure 20** makes two interesting points. First, the conversion rate is about twice as high in titanium than in aluminum; second, as discussed in Cirigliano et al. [44], different operators are beginning to produce measurable differences in the conversion rate. Hence if a signal is observed this target change would be natural. One could reasonably ask why not immediately change to very high $Z$ targets such as gold or lead, where the differences are large. The answer is straightforward: from Eckhause et al. [45], Suzuki et al. [11], or Measday [10], the muon lifetime in aluminum is 864 ns, in titanium 328 ns, and in gold only 74 ns. Since the beam pulse is $\sim$ 250 ns wide, too many of the muons would be captured and decay within the beam flash, as discussed earlier in section 2.3. Some other technology (e.g., [46]) is required: muon storage rings are a natural solution. A storage ring would eliminate the beam flash and make it possible to look for conversions much earlier in time and therefore reach higher in $Z$. Such storage rings are naturally related to rings for neutrino factories and muon colliders.

An upgrade will require the re-examination of much of the experiment. More intensity translates to more demands on the cooling and greater mechanical and thermal stress, which affect the primary proton target and the heat and radiation shield for the Production Solenoid. Increased radiation damage from more integrated intensity affects the target and shield as well. The extinction measurement will need to improve, and the time required to make a measurement will be longer. More intensity yields more dead time in the cosmic ray veto, and more integrated intensity increases the radiation damage to the readout. The tracker and calorimeter will see more damage and rate from the beam flash. The stopping target monitor technique of using germanium may not be scalable with another ×10 more intensity because of both intrinsic rate limitations or the increase in the rate of radiation damage. Such challenges are already being discussed [27].

## AUTHOR CONTRIBUTIONS

The author confirms being the sole contributor of this work and has approved it for publication.

## FUNDING

This manuscript has been authored by Fermi Research Alliance, LLC under Contract No. DE-AC02-07CH11359 with the U.S. Department of Energy, Office of Science, Office of High Energy Physics.

## ACKNOWLEDGMENTS

The author would like to thank the Mu2e Experiment for permitting the liberal use of information, and especially thank Emma Castiglia, David Hedin, and Manolis Kargiantoulakis for a thorough reading of the manuscript and many consequent improvements.

## REFERENCES

1. Calibbi L, Signorelli G. Charged lepton flavour violation: an experimental and theoretical introduction. *Riv Nuovo Cim.* (2018) **41**:71. doi: 10.1393/ncr/i2018-10144-0
2. de Gouvêa A, Vogel P. Lepton flavor and number conservation, and physics beyond the standard model. *arXiv:1303.4097* (2013). doi: 10.1016/j.ppnp.2013.03.006
3. Marciano WJ, Mori T, Roney JM. Charged lepton flavor violation experiments. *Annu Rev Nucl Part Sci.* (2008) **58**:315–41. doi: 10.1146/annurev.nucl.58.110707.171126
4. Kuno Y, Okada Y. Muon decay and physics beyond the standard model. *RevModPhys* (2001) **73**:151–202. doi: 10.1103/RevModPhys.73.151
5. Bernstein RH, Cooper PS. Charged lepton flavor violation: an experimenter's guide. *Phys Rept.* (2013) **532**:27–64. doi: 10.1016/j.physrep.2013.07.002
6. Kuno Y. Lepton flavor violation: Muon to electron conversion, COMET and PRISM/PRIME at J-PARC. In: *International Workshop on Neutrino Factories, Super Beams and Beta Beams*. Valencia (2008).
7. KInsho M, Yamamoto K, Saha PK, Ikegami M, Kawamura N, Kobayashi H. *Expression of Interest for an Experiment Searching for $\mu$-e Conversion at J-PARC Muon Facility* (2012). Available online at: http://deeme.hep.sci.osaka-u.ac.jp/documents
8. Bartoszek L, Barnes L, Miller JP, Mott A, Palladino A, Quirk J, et al. *Mu2e Technical Design Report.* FERMILAB-TM-2594, FERMILAB-DESIGN-2014-01. *arXiv:1501.05241* (2014).
9. Czarnecki A, Garcia i Tormo X, Marciano WJ. Muon decay in orbit: spectrum of high-energy electrons. *Phys Rev D* (2011) **84**:013006 doi: 10.1103/PhysRevD.84.013006
10. Measday DF. The nuclear physics of muon capture. *Phys Rep.* (2001) **354**:243–409. doi: 10.1016/S0370-1573(01)00012-6
11. Suzuki T, Measday DF, Roalsvig JP. Total nuclear capture rates for negative muons. *Phys Rev C* (1987) **35**:2212–24. doi: 10.1103/PhysRevC.35.2212
12. Feinberg G, Kabir P, Weinberg S. Transformation of Muons into electrons. *Phys Rev Lett.* (1959) **3**:527–30. doi: 10.1103/PhysRevLett.3.527
13. Berryman JM, de Gouveâ A, Kelly KJ, Kobach A. Lepton-number-violating searches for muon to positron conversion. *Phys Rev.* (2017) **D95**:115010. doi: 10.1103/PhysRevD.95.115010
14. Geib T, Merle A, Zuber K. $\mu^- - e^+$ conversion in upcoming LFV experiments. *Phys Lett.* (2017) **B764**:157–62. doi: 10.1016/j.physletb.2016.11.029
15. Commins ED. *Weak Interactions.* New York, NY: Cambridge University Press; Mc-Graw Hill, Inc. (1973).
16. Michel L. Interaction between four half-spin particles and the decay of the $\mu$-meson. *Proc Phys Soc A* (1950) **63**:514.
17. Perkins DH. *Introduction to High Energy Physics.* 4th Edn. New York, NY: Cambridge University Press (2000).
18. Czarnecki A, Dowling M, Garcia i Tormo X, Marciano WJ, Szafron R. Michel decay spectrum for a muon bound to a nucleus. *Phys Rev.* (2014) **D90**:093002. doi: 10.1103/PhysRevD.90.093002
19. Bergbusch PC, Armstrong DS, Blecher M, Chen CQ, Doyle BC, Gorringe TP, et al. Radiative muon capture on O, Al, Si, Ti, Zr, and Ag. *Phys Rev C* (1999) **59**:2853–64. doi: 10.1103/PhysRevC.59.2853
20. Kroll NM, Wada W. Internal pair production associated with the emission of high-energy gamma rays. *Phys Rev.* (1955) **98**:1355–9. doi: 10.1103/PhysRev.98.1355
21. Bertl W, Engfer R, Hermes EA, Kurz G, Kozlowski T, Kuth J, et al. A search for $\mu$-e conversion in muonic gold. *Eur Phys J C* (2006) **47**:337–46. doi: 10.1140/epjc/s2006-02582-x
22. Christillin P, Rosa-Clot M, Servadio S. Radiative muon capture in medium-heavy nuclei. *Nucl Phys A* (1980) **345**:331–66.
23. Jackson JD. *Classical Electrodynamics.* New York, NY: Cambridge University Press; John Wiley & Sons (1975).
24. *Mathematica. Version 11.3.* Champaign, IL: Wolfram Research, Inc. (2018).
25. di Mauro M, Donato F, Goudelis A, Serpico PD. New evaluation of the antiproton production cross section for cosmic ray studies. *Phys Rev D* (2014) **90**:085017. doi: 10.1103/PhysRevD.90.085017
26. Boyarinov SV, Evseev II, Kiselev YT, Leksin GA, Martem'yanov AN, Mikhailov KR, et al. Yields of $p, \bar{p}, \pi^{\pm}$, and $K^{\pm}$ emitted at an angle of $97^o$ in the laboratory system from nuclei irradiated by 10.14-GeV protons. *Phys Atom Nuclei* (1994) **57**:1452–62.
27. Abusalma F, Ambrose D, Artikov A, Bernstein R, Blazey GC, Bloise C, et al. Expression of interest for evolution of the Mu2e experiment, FERMILAB-FN-1052. *arXiv:1802.02599* (2018).
28. Abadjev VS, Bakhtin BN, Goncharenko ON, Djilkibaev RM, Edlichka VV, Lobashev VM. *MELC Experiment to Search for the $\mu$A → eA Process.* INR–786/92 (1992). Available online at: https://mu2e-docdb.fnal.gov:440/cgi-bin/RetrieveFile?docid=76;filename=meco002.pdf;version=1
29. Stekly ZJJ, Zar JL. Stable superconducting coils. *IEEE Trans Nucl Sci.* (1965) **12**:367–72.
30. Tanabashi M, Hagiwara K, Hikasa K, Nakamura K, Sumino Y, Takahashi F, et al. Review of particle physics. *Chin Phys.* (2016) **C40**:100001. doi: 10.1103/PhysRevD.98.030001
31. Roberts TJ, Beard KB, Ahmed S, Huang D, Kaplan DM. G4beamline particle tracking in matter dominated beam lines. *Conf Proc.* (2011) **C110328**:373–5.
32. MacLachlan J. Particle tracking in E-Phi space for synchrotron design & diagnosis. In: *Presented at Conference: C92-11-02 Proceedings FERMILAB-CONF-92-333* (1992). Available online at: http://esme.fnal.gov
33. Kosmas TS, Vergados JD, Faessler A. Muon number violating processes in nuclei. *Phys Atom Nucl.* (1998) **61**:1161–74.
34. Knoll GF. *Radiation Detection and Measurement*. 4th Edn. New York, NY: Wiley (2010).
35. Miyashita T. The mu2e experiment. In: *Presentation at 51st Annual Fermilab Users Meeting* (2018). Available online at: https://indico.fnal.gov/event/16332/session/5/contribution/17/material/slides/0.pdf

36. Bonventre R. Tollestrup prize lecture. In: *Presentation at 51st Annual Fermilab Users Meeting* (2018). Available online at: https://indico.fnal.gov/event/16332/session/5/contribution/18/material/slides/0.pdf
37. Lucà A. A panel prototype for the Mu2e straw tube tracker at fermilab. In: *Proceedings, Meeting of the APS Division of Particles and Fields (DPF 2017): Fermilab*. Batavia, IL (2017). Available online at: https://inspirehep.net/record/1629960/files/arXiv:1710.03799.pdf
38. Atanov N, Baranov V, Budagov J, Cervelli F, Colao F, Cordelli M, et al. The Mu2e undoped CsI crystal calorimeter. *JINST* (2018) **13**:C02037. doi: 10.1088/1748-0221/13/02/C02037
39. Artikov A, Baranov V, Blazey GC, Chen N, Chokheli D, Davydov Y, et al. Photoelectron yields of scintillation counters with embedded wavelength-shifting fibers read out with silicon photomultipliers. *Nucl Instrum Methods* (2018) **A890**:84–95. doi: 10.1016/j.nima.2018.02.023
40. Dukes EC, Ehrlich R. A high efficiency cosmic ray veto for the Mu2e experiment. In: *Conference: 38th International Conference on High Energy Physics* New York, NY: Cambridge University Press (2016).
41. Feldman GJ, Cousins RD. Unified approach to the classical statistical analysis of small signals. *Phys Rev D* (1998) **57**:3873–89. doi: 10.1103/PhysRevD.57.3873
42. Neuffer D. *Mu2e-II Injection From PIP-II*. Fermi National Accelerator Lab, Batavia, IL (2018).
43. Ball M, Burov A, Chase B, Chakravarty A, Chen A, Dixon S, et al. *The PIP-II Conceptual Design Report*. Argonne National Lab. (ANL), Argonne, IL; Fermi National Accelerator Lab. (FNAL), Batavia, IL (2017).
44. Cirigliano V, Kitano R, Okada Y, Tuzon P. Model discriminating power of $\mu \to e$ conversion in nuclei. *Phys Rev D* (2009) **80**:013002. doi: 10.1103/PhysRevD.80.013002
45. Eckhause M, Siegel RT, Welsh RE, Filippas TA. Muon capture rates in complex nuclei. *Nucl Phys.* (1966) **81**:575–84.
46. Kuno Y. PRISM/PRIME. *Nucl Phys B* (2005) **149**:376–8. doi: 10.1016/j.nuclphysbps.2005.05.073

**Conflict of Interest Statement:** The author declares that the research was conducted in the absence of any commercial or financial relationships that could be construed as a potential conflict of interest.